\documentclass[prb,reprint]{revtex4-2}
\usepackage{graphicx,amssymb,amsfonts,amsmath}
\usepackage{color}
\usepackage{mathtools}
\usepackage{latexsym}
\usepackage{graphicx}
\usepackage{float}
\usepackage{dcolumn}
\usepackage{bm}
\usepackage{amssymb}
\usepackage{amsmath}
\usepackage{mathtools}
\usepackage[space]{grffile}
\usepackage{xcolor}
\usepackage{comment}
\usepackage{ulem}

\usepackage{todonotes}

\usepackage{hyperref}
\hypersetup{
    colorlinks=true,
    linkcolor=blue,
    filecolor=blue,      
    urlcolor=blue,
    citecolor=blue
    }

\begin{document}

\title{Slowly decaying zero mode in a weakly non-integrable boundary impurity model}
\author{Hsiu-Chung Yeh$^{1}$}
\author{Gabriel Cardoso$^{2}$}
\author{Leonid Korneev$^{3}$}
\author{Dries Sels$^{1,4}$}
\author{Alexander G. Abanov$^{3,5,6}$}
\author{Aditi Mitra$^{1}$}
\affiliation{
$^{1}$Center for Quantum Phenomena, Department of Physics,
New York University, 726 Broadway, New York, NY, 10003, USA\\
$^{2}$Tsung-Dao Lee Institute, Shanghai Jiao Tong University, Shanghai, 201210, China\\
$^{3}$Department of Physics and Astronomy, Stony Brook University, Stony Brook, NY 11794, USA\\
$^{4}$Center for Computational Quantum Physics, Flatiron Institute, New York, New York 10010, USA\\
$^{5}$Simons Center for Geometry and Physics, Stony Brook, NY 11794, USA \\
$^{6}$The Rosi and Max Varon Visiting Professor/Fellow, Weizmann Institute of Science, Israel
}

\begin{abstract}
The transverse field Ising model (TFIM) on the half-infinite chain possesses an edge zero mode. This work considers an impurity model --- TFIM perturbed by a boundary integrability breaking interaction. For sufficiently large transverse field, but in the ordered phase of the TFIM, the zero mode is observed to decay. The decay is qualitatively different from zero modes where the integrability breaking interactions are non-zero all along the chain.  It is shown that for the impurity model, the zero mode decays by relaxing to a non-local quasi-conserved operator, the latter being exactly conserved when the opposite edge of the chain has no non-commuting perturbations so as to ensure perfect degeneracy of the spectrum. In the thermodynamic limit, the quasi-conserved operator vanishes, and a regime is identified where the decay of the zero mode obeys Fermi's Golden Rule. A toy model for the decay is constructed in Krylov space and it is highlighted how Fermi's Golden Rule may be recovered from this toy model. 
\end{abstract}
\maketitle

\section{Introduction}

The transverse field Ising model (TFIM) with open boundary conditions hosts Majorana zero modes  \cite{Kitaev01}. These zero modes are also known as strong zero modes where the edge mode is associated with an operator that commutes with the Hamiltonian
in the thermodynamic limit and anti-commutes with a discrete $Z_2$ symmetry \cite{Fendley2012,FendleyXYZ,Sen13,Alicea16,Fendley17,Nayak17,Garrahan18,Garrahan19,Yates19,Yates20a,Yao20,Yates20,Yates21,Yates22,Bardarson22,Fendley23,yeh2023decay}. Thus the existence of a strong zero mode implies a doubly degenerate spectrum, with an observation of the zero mode not tied to the ground state sector. How perturbing away from the TFIM affects the strong zero mode is an essential question, as it concerns practical applications where the experimental set-up is only approximately a TFIM. Understanding this is also important from a conceptual point of view as it addresses how thermalization times are affected by quasi-conserved quantities.  

After the Jordan-Wigner transformation, the TFIM maps to a 1D model of spinless fermions with nearest-neighbor hopping, with the $Z_2$ symmetry corresponding to fermion parity. A standard way to perturb away from this model is to include four fermion interactions \cite{Fendley17,Nayak17,Yates19,Yates20,Yao20,Yates20a,Yates21,Yates22,yeh2023decay}. Here we study numerically the effect of a weaker perturbation, where the four fermion interactions exist only at the boundary. This work aims to explore whether such a  boundary integrability breaking term can destroy the zero mode, and if so, what is the signature of the decay of the zero mode in the dynamics. The boundary perturbation we consider is not equivalent to the family of integrability preserving boundary conditions of the TFIM \cite{ghoshal1994boundary}. It breaks the integrability of the spin chain. The Majorana excitations incident at the boundary might be reflected as a single Majorana or a triplet of Majoranas, according to the field theory model considered in Ref.~\cite{arthur2016breaking}. Our focus here is the effect of the boundary integrability breaking on the existence of the strong zero mode.  

It is not easy to establish numerically whether boundary perturbations can cause the zero mode to decay. This is because, even for the TFIM, where a zero mode can be analytically constructed, a finite system size $L$ causes the zero mode to decay. This decay comes about because of tunneling processes that hybridize the zero modes at the two ends of the chain, leading to a lifetime that is exponential in the system size. In the presence of perturbations, typically, one needs to park oneself at some parameter regime where the decay becomes $L$-independent, and only then one can safely claim that the  boundary perturbation destroys the zero mode.

In this paper, we study the system using a combination of three different methods, (i) exact diagonalization (up to system sizes of $L=14$), (ii) Trotterized time-evolution of Haar random states (up to $L=22$ and $t=10^4/J_x$, $J_x$ being the strength of the Ising interaction in the TFIM), which approximates the real dynamics up to exponential times and with exponential precision in space, (iii) Krylov space dynamics, which allows us to construct an approximate model for the zero mode decay in the thermodynamic limit.

The paper is organized as follows. In Section \ref{sec:Model} we outline the model and explain how the zero mode is detected numerically.
In Section \ref{sec:Quasi-conserved quantity}, we construct a quasi-conserved quantity, which becomes exactly conserved when non-commuting couplings at one end of the chain are switched off. We highlight the role that the quasi-conserved quantity plays in the decay of the zero mode. In Section \ref{sec:Perturbation theory of decay rate}, we park ourselves in a region of parameter space where the decay is entirely due to processes that are second order in the integrability breaking term, deriving the Fermi Golden Rule (FGR) decay rate, and comparing it with numerics. We present our conclusions in Section \ref{sec:Conc}.
In Appendix \ref{Appendix:Krylov} we outline how the zero mode can be studied using Krylov space methods. We derive an effective model for the zero mode decay and highlight how FGR is recovered in  Krylov sub-space. In Appendix \ref{Appendix:Quasi-conserved operator spin chain}, we present examples of the quasi-conserved quantities in short spin chains. In Appendix \ref{Appendix:Random state and Trotter}, we outline the numerical method of Haar random average and judicious Trotter decomposition, while in Appendix \ref{Appendix:FGR} we provide details in the derivation of the FGR decay rate.

\section{Model} \label{sec:Model}

We study the TFIM of length $L$ with open boundary conditions and perturbed by a boundary impurity. The latter is
modeled as an integrability-breaking exchange interaction acting only on the first two sites of the chain. Thus the Hamiltonian is 
\begin{align}
    H =  J_x \sum_{i=1}^{L-1} \sigma_i^x \sigma_{i+1}^x +g \sum_{i=1}^{L} \sigma_i^z + J_z \sigma_1^z\sigma_2^z,
    \label{Eq: Impurity Hamiltonian}
\end{align}
where $\sigma^{x,y,z}_i$ are Pauli matrices on site $i$,  $g$ is the strength of transverse field and $J_{x,z}$ is the strength of the Ising interaction in the $x,z$-directions, with $J_z$ being non-zero only on the first link. We will set $J_x = 1$ in the paper.

For $J_z$=$0$, the Hamiltonian is the TFIM, $H_0 = H|_{J_z = 0}$, which is in the topological phase with an edge zero mode 
for $|g| < 1$. The zero mode operator $\psi_0$ anti-commutes with the $Z_2$ symmetry, $\mathcal{D} = \sigma_1^z \ldots \sigma_L^z$ of the system: $\{ \psi_0, \mathcal{D} \} = 0$. In the thermodynamic limit of a semi-infinite chain, the zero mode commutes with the TFIM, $[\psi_0 ,H_0] = 0$. Because of this property, the edge zero mode has an infinite lifetime in the thermodynamic limit.

On adding integrability-breaking perturbations, the commutation relation between the zero mode and the Hamiltonian no longer holds. However, one can still observe a long-lived quasi-stable edge zero mode for  boundary integrability breaking. 
A useful quantity to probe this object is the infinite temperature autocorrelation of $\sigma_1^x$
\begin{align}
    A_\infty (t) = \frac{1}{2^L} \text{Tr}[\sigma_1^x(t) \sigma_1^x],
    \label{Eq: AutoCorrelation Sigmax}
\end{align}
where $t$ is the time measured in units of $1/J_x$. This is a good measure of the zero mode lifetime in the presence of interactions since the zero mode is localized on the edge with $\mathcal{O}(1)$ overlap with $\sigma_1^x$, $\text{Tr}[\psi_0 \sigma_1^x]/2^L \sim \mathcal{O}(1)$ \cite{Kitaev01,FendleyXYZ}. In the language of Majorana fermions, $\sigma_1^x$ is the Majorana fermion on the first site, and the edge mode is a superposition of Majoranas, with the largest weight being on the Majoranas on the first few sites at the boundary. 

\begin{figure}[h]
    \centering \includegraphics[width=0.45\textwidth]{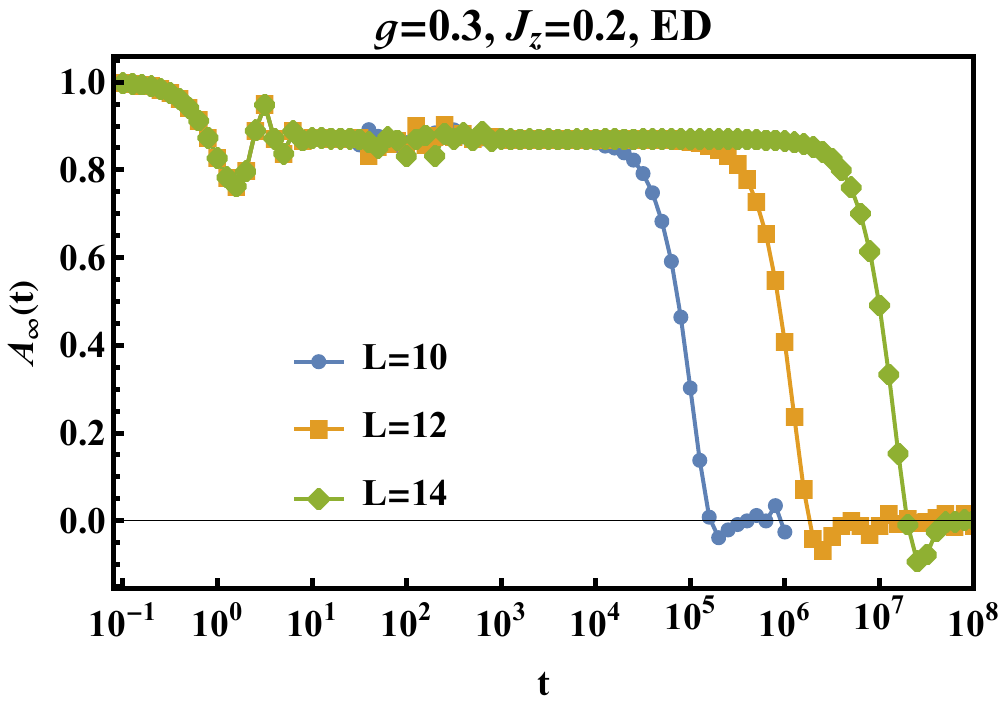} \includegraphics[width=0.45\textwidth]{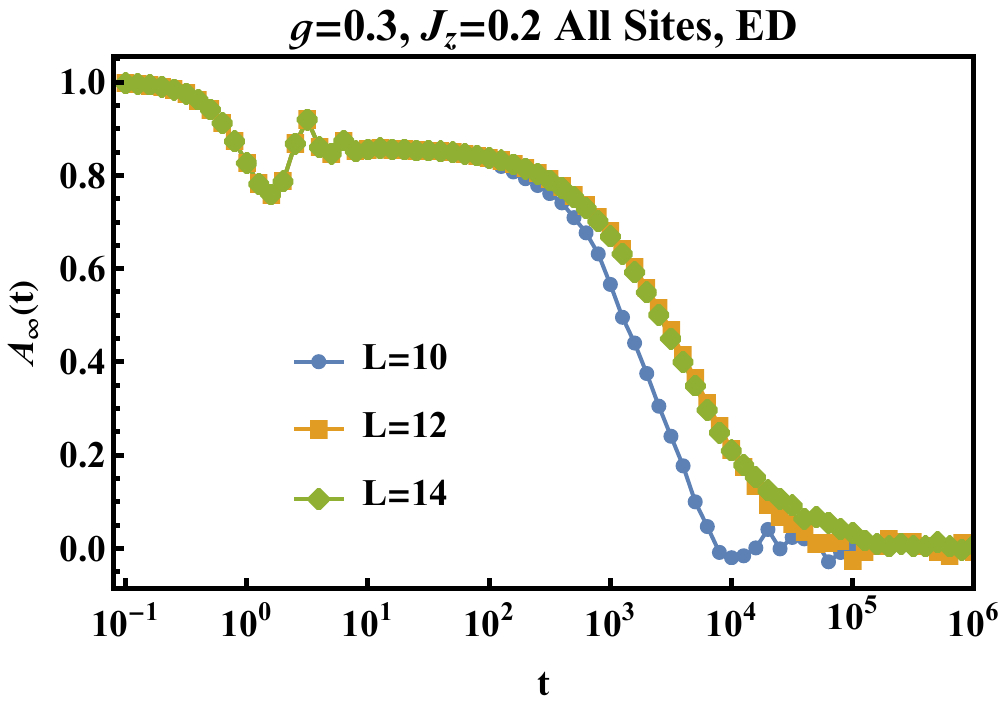}
    \caption{Infinite temperature autocorrelation function  for boundary impurity  (top panel) and perturbation on all sites  (bottom panel) with $g=0.3$ and $J_z =0.2$. For the boundary impurity (top panel), the edge zero mode survives for long times and does not show saturation of lifetime with system sizes up to $L=14$. The model with perturbations on  all sites (bottom panel) also shows a long-lived edge zero mode, but with a shorter lifetime that has saturated at $L=12$.}
    \label{fig: autocorrelation g=0.3}
\end{figure}

Fig.~\ref{fig: autocorrelation g=0.3}  shows examples of the autocorrelation function for $g=0.3$ and two different models. The top panel shows autocorrelation functions for the boundary impurity \eqref{Eq: Impurity Hamiltonian}. For comparison, the bottom panel presents the autocorrelation functions of the chain with a $J_z$ perturbation on all sites, $J_z \sum_{i=1}^{L-1} \sigma_i^z \sigma_{i+1}^z$.
The bottom panel shows that after an initial transient, the autocorrelation decays into a long-lived zero mode which lasts for a long time, as shown by the constant value of the autocorrelation function. The overlap of the plots for $L=12,14$ in the bottom panel suggests that the eventual decay of the autocorrelation to zero is due to interactions rather than finite system size.  

In contrast, the boundary impurity model (top panel) shows a much longer lifetime due to the weaker nature of the integrability breaking perturbation ($J_z$ non-zero only on the first link). In particular, the autocorrelation does not show saturation of lifetime when system size increases up to size $L=14$ in contrast to the bottom panel. The top panel seemingly suggests an exact zero mode instead of a quasi-stable zero mode for the impurity model. 
However, this appears to be a  finite system size effect, and that for the given parameters, we simply do not have access to large enough $L$ to be in a regime where the decay is dominated by interactions.  This is supported by the fact that as one increases the transverse field, the autocorrelation shows a tendency to saturate with increasing system size. But, interestingly, even in this regime of eventual $L$-independent decay, there is still a qualitative difference in the decay mechanism of the zero mode for the impurity model and that for the model where $J_z\neq0$ on all links.  

\begin{figure}[h]
    \centering \includegraphics[width=0.45\textwidth]{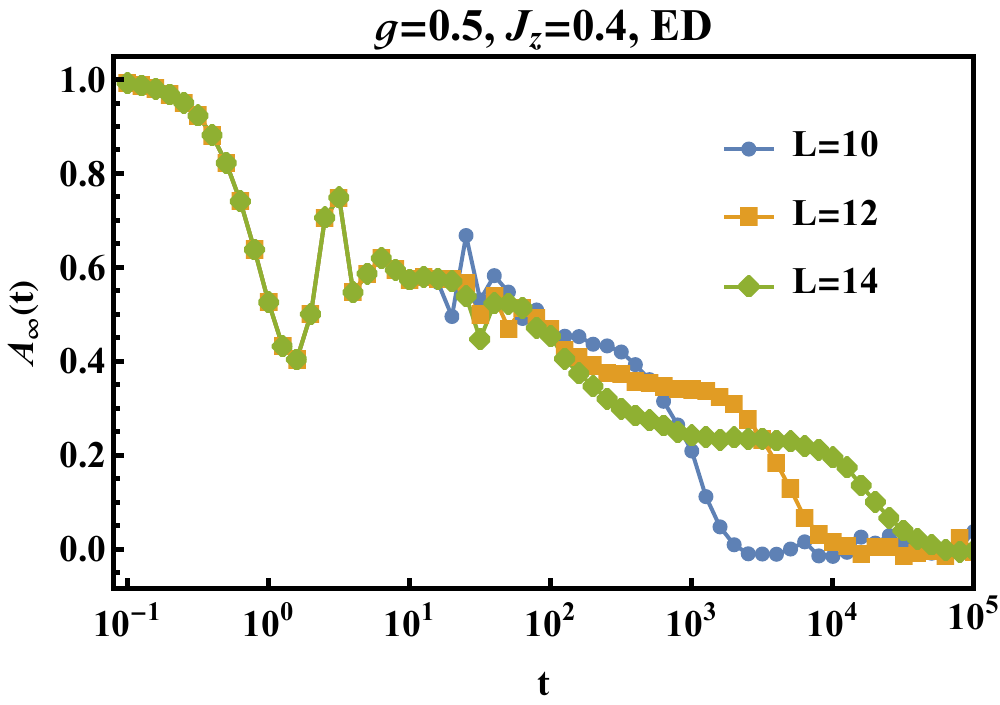}
    \caption{Infinite temperature autocorrelation function for the boundary impurity with $g=0.5$ and $J_z = 0.4$. Unlike Fig.~\ref{fig: autocorrelation g=0.3} which was for a smaller $g,J_z$, the autocorrelation shows a plateau region before decaying to zero (see, e.g., the range of $t$ between $10^3$ and $10^4$ for the $L=14$ plot). The height of the plateau decreases as the system size increases.}
    \label{fig: autocorrelation g=0.5}
\end{figure}

In Fig.~\ref{fig: autocorrelation g=0.5} we identify three steps in the decay of the autocorrelation function for a finite-size impurity model: (i) after an initial transient (see $t<10^1$) the autocorrelation decays into  the local zero mode of the original non-perturbed Hamiltonian $H_0$  (see $t <10^2$ for $L=14$), (ii) the system decays from this local zero mode to another quasi-conserved operator, reflected by a second plateau
from $t=10^3-10^4$ for $L=14$, (iii) it finally reaches zero due to the interaction  ($t>10^4$ for $L=14$). 
The plateau value at the end of step (ii) decreases with system size, indicating the existence of some non-local quasi-conserved operator. This effect was not observed for the model with perturbations on all sites, a fact which will be highlighted further later. 
In the thermodynamic limit, the plateau value goes to zero, and step (iii) disappears eventually. We expect that in the thermodynamic limit, the decay rate is dominated by step (ii), and we will show that, in certain regimes, this decay can be captured by perturbation theory in $J_z$.

In the following section, we will demonstrate the existence of the non-local quasi-conserved operator and highlight its role in the decay of the zero mode.

\section{Quasi-conserved operator} \label{sec:Quasi-conserved quantity}

To construct the quasi-conserved operator, we follow the argument by Fendley \cite{FendleyXYZ} on the commutation relation between the zero mode and the Hamiltonian. The zero mode of an integrable model such as the TFIM \cite{FendleyXYZ}, $XY$ chain \cite{Yates20a} or XYZ chain \cite{FendleyXYZ}, does not commute with the integrable Hamiltonian at any finite system size, but only in the thermodynamic limit. However, for finite system size, the zero mode commutes with almost the whole Hamiltonian except for the interaction terms on the last site \cite{FendleyXYZ,Yates20a}. For example, for the TFIM
\begin{align}
    H_0 = \sum_{i=1}^{L-1} \sigma_i^x \sigma_{i+1}^x + g \sum_{i=1}^{L} \sigma_i^z,
    \label{Eq: Transverse-field Ising}
\end{align}
the corresponding zero mode localized on the first site is given by the following superposition of Majoranas up to an overall normalization
\begin{align}
    \psi_0 \propto \sum_{l=1}^{L} g^{l-1} a_{2l-1},
    \label{Eq: Zero mode of Transverse-field Ising}
\end{align}
where the Majoranas are defined as follows
\begin{align}
    &a_{2l-1} = \prod_{j=1}^{l-1}\sigma_j^z \sigma_l^x; & a_{2l} = \prod_{j=1}^{l-1}\sigma_j^z \sigma_l^y.
    \label{Eq: Majorana}
\end{align}
The commutation between the zero mode and the Hamiltonian is non-zero due to the transverse field on the last site, $[\psi_0 ,H_0] = [\psi_0 ,g\sigma_L^z] \neq 0$. However, since the zero mode is localized on the first site, this commutation is exponentially small in $L$ and becomes zero in the thermodynamic limit.

Based on the above argument, one can numerically construct a conserved operator $O_c$ in the following way. Let us take the TFIM as an example. 
Consider first the TFIM with the last-site transverse field turned off, $\Tilde{H}_0 = H_0 - g\sigma_L^z$.
Now, $\psi_0$ exactly commutes with $\tilde{H}_0$, as does $\sigma^x_L$, and these two operators commute with each other $\left[\psi_0,\sigma^x_L\right]=0$. The Hamiltonian $\Tilde{H}_0$ has an exactly two-fold degenerate energy spectrum for any finite system size. In particular, $\Tilde{H}_0$ splits into two sectors labeled by the eigenstates of parity $\mathcal{D}$, but with both $\sigma^x_L$ and $\psi_0$
flipping between the states of two different parities. Since $\sigma^x_L,\psi_0$ precisely commute with $\tilde{H}_0$, this ensures an exact double degeneracy.

Given such a double degeneracy of the spectrum, one may construct a conserved operator which is odd under $Z_2$, and has non-zero matrix elements between opposite parity eigenstates of $\Tilde{H}_0$. In addition, one may choose this operator to overlap with
$\sigma^x_1$. Such a conserved operator $O_c$, is the long time-limit of the operator $\sigma_1^x(t)$ (up to an overall normalization) 
\begin{align}
    O_c \propto \lim_{T \rightarrow \infty} \frac{1}{T} \int_0^T dt\ \sigma_1^x(t).
    \label{Eq: O from time average}
\end{align}
Numerically, the conserved operator $O_c$ can be constructed by eliminating the terms oscillating in $\sigma_1^x(t)$. In the eigenbasis of $\tilde{H}_0$, $\sigma_1^x(t)$ is represented by
\begin{align}
    \langle \tilde{n}| \sigma_1^x (t) |\tilde{m} \rangle = \langle \tilde{n}| \sigma_1^x |\tilde{m} \rangle e^{i(\tilde{E}_n - \tilde{E}_m)t},
\end{align}
where the matrix elements are not oscillating as long as $\tilde{E}_n = \tilde{E}_m$. In numerics, one is only required to construct matrix elements with $\tilde{E}_n = \tilde{E}_m$, 
\begin{align}
   \langle \tilde{n}| O_c |\tilde{m} \rangle = \Big\{\begin{array}{cc}
    \langle \tilde{n}| \sigma_1^x |\tilde{m} \rangle & \text{if}\ \tilde{E}_n = \tilde{E}_m;\\
     0 & \text{else}.
\end{array}
\label{Eq: O from ED}
\end{align}
Finally, $O_c$ is normalized with a norm equal to one, $\text{Tr}[O_c^\dagger O_c]/2^L = 1$.  In the example of the TFIM, the conserved operator $O_c$ happens to be the same as the zero mode $\psi_0$ \eqref{Eq: Zero mode of Transverse-field Ising}. Still, they may be different in generic models possessing zero modes. For constructing the zero mode in generic models, it is proposed to apply commutant algebra, and this may be achieved both analytically and numerically, see \cite{moudgalya2023symmetries,moudgalya2023numerical}.

The physical reason behind the conserved quantity $O_c$ is that for the TFIM, there are two Majorana modes, one on the left end, 
and the other on the right end of the chain. The
lifetime comes from the two modes coupling via tunneling processes, with the tunneling amplitude $\propto g^L$. However, when $g$ is made zero on the last site, the two zero modes,
one related to $\sigma^x_1$, and the other related to $\sigma^x_L$, no longer hybridize. Thus
the zero mode on the left end does not decay and is exactly conserved.

\begin{figure}[h]
    \centering
    \includegraphics[width=0.45\textwidth]{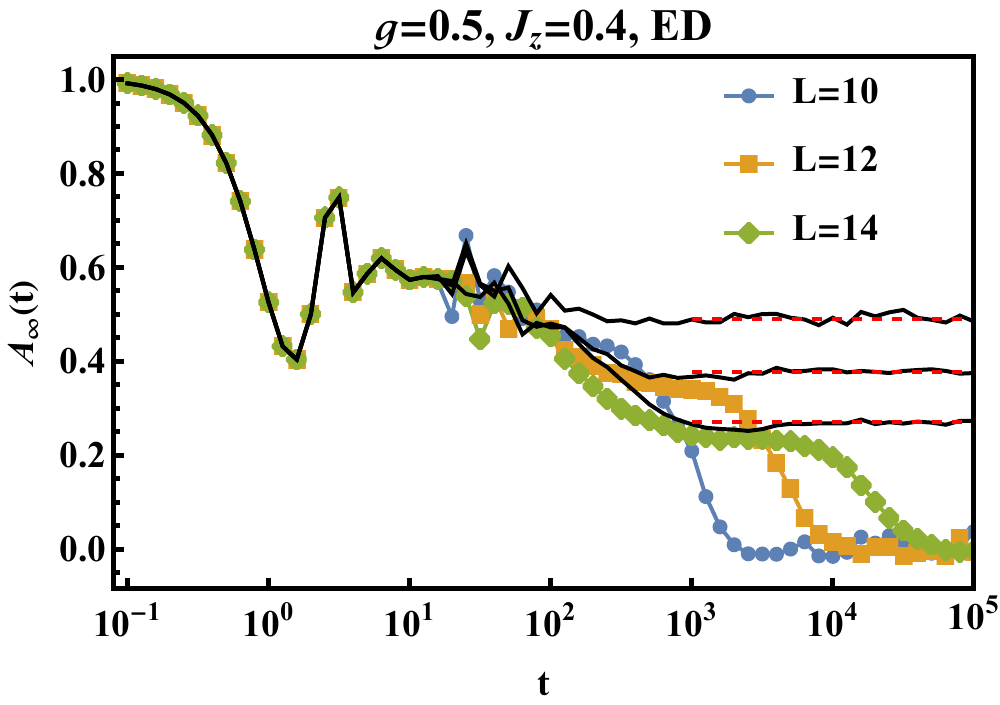}
    \includegraphics[width=0.45\textwidth]{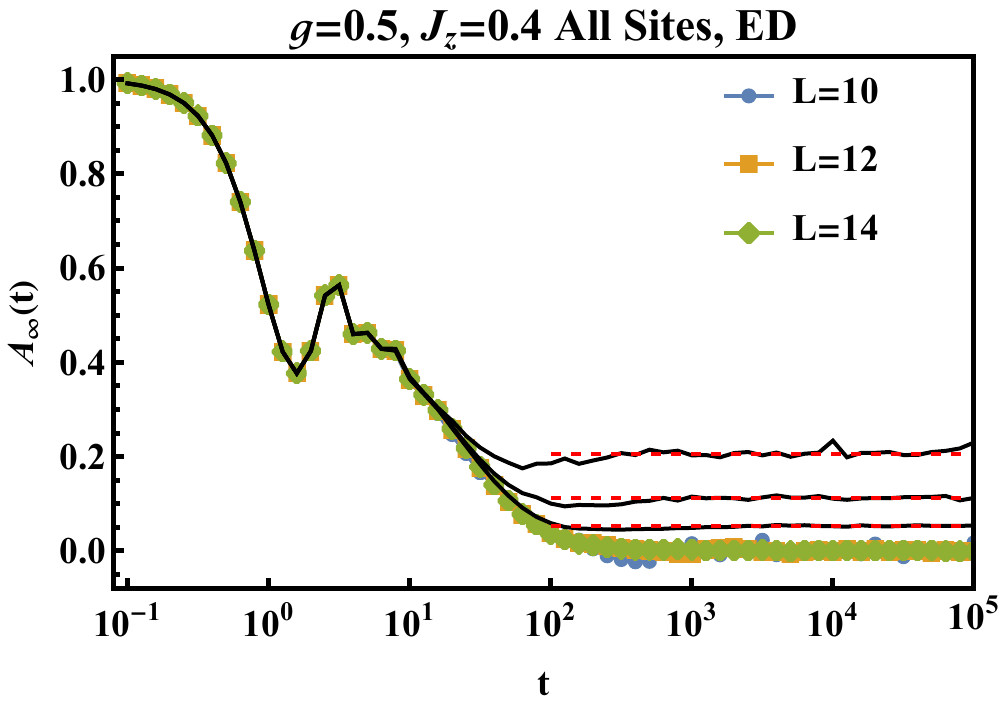}
    \caption{Infinite temperature autocorrelation function for the boundary impurity model (top panel) and  for model with perturbations on all sites (bottom panel) with $g=0.5$ and $J_z =0.4$.  The autocorrelation with interactions switched off on the last site (i.e., $g=0$ on the last site in the top panel and $g=0,J_z=0$ on the last site in the bottom panel) are plotted as well (solid black lines) and show the existence of a conserved operator at late times. This is consistent with the numerically constructed zero mode in \eqref{Eq: O from ED}, highlighted with red dashed line for late times. When the interactions on the last site are restored, the conserved operator immediately disappears for the case of all-site perturbation (bottom panel) but becomes quasi-stable for the boundary impurity model (top panel). The plateaus in the top panel  are the remnant of the conserved operator.}
    \label{fig: autocorrelation and conserved case}
\end{figure}

In general, the method of constructing the conserved operator $O_c$ described above can be applied to any spin system as long as the two-fold degeneracy of the energy spectrum can be achieved by turning off interactions on the last site. Moreover, one should choose an appropriate seed operator to numerically generate the conserved operator; e.g., we choose $\sigma_1^x$ as the seed operator since $J_x=1$ is the largest coupling, and $\sigma_1^x$ is the first Majorana in the convention \eqref{Eq: Majorana}. However, once the interaction on the last site is restored, the conserved operator may become quasi-conserved, or may even immediately die out, if the commutation with the interactions on the last site does not approach zero in the thermodynamic limit. In Fig.~\ref{fig: autocorrelation and conserved case}, we demonstrate the autocorrelation with $g = 0.5, J_z = 0.4$ for both cases: the boundary impurity model and the model with non-zero perturbations on all sites. Also plotted are the autocorrelations with no interactions on the last site except $\sigma_{L-1}^x\sigma_{L}^x$. For comparison, the numerically constructed zero mode from \eqref{Eq: O from ED} are also plotted as red dashed lines. The agreement between the plateaus (solid black lines) and this numerically constructed zero mode (dashed red lines) is excellent.

In the case of perturbations on all sites (bottom panel in Fig.~\ref{fig: autocorrelation and conserved case}), the autocorrelation already saturates at small system sizes $L = 10$. When the interaction is turned off on the last site, the autocorrelation approaches a non-zero constant value (solid black lines) which corresponds to the conserved operator \eqref{Eq: O from time average}. The decrease of the plateau height with increasing system size indicates that the conserved operator becomes less localized on the first Majorana. Note that this plateau formation happens at times longer than the decay time for the model where $J_z$ is non-zero along the chain.
As the interactions on the last site are restored, the conserved operator immediately dies out due to interactions.

In contrast, for the impurity model where the perturbation is present only at the boundary (top panel in Fig.~\ref{fig: autocorrelation and conserved case}), it is clearly seen that the late-time plateau comes from the conserved operator. Once the last-site interaction is restored, the operator becomes quasi-conversed so that the autocorrelation persists for a long time before it eventually decays. The quasi-conserved operator is non-local as shown by the decrease of the plateau value as $L$ increases, going to zero in the thermodynamic limit.

The different behaviors of the autocorrelation between the two models, perturbations on all sites, and the boundary impurity model are due to the different bulk properties of the two models; the former is non-integrable, while the latter is free in the bulk. One can also capture the difference between the two autocorrelation functions by mapping the dynamics of $\sigma^x_1$ to  
single-particle dynamics in Krylov space. The Krylov space Hamiltonian is a tri-diagonal Hamiltonian, where the off-diagonal elements have some universal features \cite{parker2019universal} that clearly distinguish between the two models. Moreover, decay rates can be derived by further coarse-graining the Krylov Hamiltonian and mapping it to a Dirac model with a spatially inhomogeneous mass (in Krylov space), see discussion in Appendix \ref{Appendix:Krylov}.

Note that we consider $O_c$ as a time average of $\sigma_1^x$ in the discussion. In principle, one can construct $O_c$ 
by solving the commutation relation $[\Tilde{H},O_c] = 0$ directly. However, this is only feasible for small system sizes. In Appendix~\ref{Appendix:Quasi-conserved operator spin chain}, we present analytic solutions for small system sizes and also show the equivalence between $[\Tilde{H},O_c] = 0$ and the time averaged method.

Here we summarize the physical picture for the boundary impurity model to emphasize the three steps involved in the decay of the  zero mode.
(i) The autocorrelation decays into the local zero mode $\psi_0$ of the original non-perturbed Hamiltonian $H_0$ after an initial transient. This accounts for the presence of the ``first plateau'' ($t \sim 10$ in top panel in Fig.~\ref{fig: autocorrelation and conserved case}). (ii) The local zero mode $\psi_0$ decays to another quasi-conserved operator $O_c$. This corresponds to the transition to the plateau at late times ($t \sim 10^3$ for $L=14$ in the top panel in Fig.~\ref{fig: autocorrelation and conserved case}). This plateau is a finite system size effect because the plateau height decreases as $L$ increases. (iii) The autocorrelation finally decays to zero due to interactions. This paves the way to the next section, where we focus on the decay in step (ii), i.e, the decay from $\psi_0$ to $O_c$, where the existence of $O_c$ is a finite system size effect. The step (ii) decay becomes the decay of the zero mode
in the thermodynamic limit, as in this limit the plateau due to $O_c$ vanishes.

\section{Fermi's Golden Rule decay rate} \label{sec:Perturbation theory of decay rate}

This section considers sufficiently large transverse fields where one can obtain system-size independent results. Moreover, this choice places us in a regime where Fermi's Golden Rule (FGR) approximation for the decay rate is valid. We will derive and compare the FGR decay rate with numerics.

Let us start by presenting a numerical method that allows us to compute the autocorrelation for system sizes beyond $L=14$.

Due to the limitations of computational resources, ED can only be applied up to $L = 14$. Therefore, numerically approximate methods for computing the autocorrelation are required for accessing larger system sizes. Here we outline one such approximation.
First, one approximates the trace by the average of a Haar random state $\phi$: $\text{Tr}[\cdots]/2^L \approx \langle\phi| \cdots |\phi\rangle$. The average of the Haar random state consists of two parts: diagonal and off-diagonal matrix elements in the eigenbasis representation (see Appendix \ref{Appendix:Random state and Trotter}). The diagonal part corresponds to the trace that one wants to compute. The sum of the off-diagonal parts is essentially a summation of random numbers, which is typically $\sim 1/\sqrt{2^L}$ and negligible as long as the system size is large, and the sum of diagonal parts is an $\mathcal{O}(1)$ number. Therefore, one can calculate autocorrelations up to $\mathcal{O}(1/\sqrt{2^L})$ precision without performing ED. Second, the unitary evolution is approximated by Trotter decomposition with finite time step $dt$: $U(dt) \approx \exp(-iH_{\rm xx} dt) \exp(-iH_z dt)\exp(-iH_{\rm zz} dt)$, where $H_{\rm xx}, H_z$ and $H_{\rm zz}$ correspond to the three parts of the Hamiltonian \eqref{Eq: Impurity Hamiltonian}. Physically, we have replaced the continuous time evolution with a discrete-time (Floquet) one. One recovers continuous-time dynamics in the high-frequency limit, $dt \ll 1$. Setting $dt = 0.2$, the heating time of such a Floquet system is estimated to be $\sim e^{2\pi/dt} \sim 10^{13}$.
Here, we choose $g = 0.6$ so that the autocorrelation almost decays by $t = 10^4$ for $J_z$ between $0.15-0.5$, while at the same time, this time scale is much smaller than the heating time.  Combining these two approximations, the autocorrelation can be massaged into the average of a Haar random state at different times. Computationally, one only requires to perform the time evolution of a state. This costs significantly less resources than ED so that one can probe larger system sizes. However, it is inefficient for calculating long-time behavior since the computation time is proportional to the number of time steps fixed by the Trotter decomposition; see details of numerical methods and discussion in Appendix \ref{Appendix:Random state and Trotter}. This is the main reason why  a larger transverse field strength $g=0.6$ is chosen, allowing us to study system sizes up to $L=22$. 

\begin{figure}[h]
    \centering \includegraphics[width=0.45\textwidth]{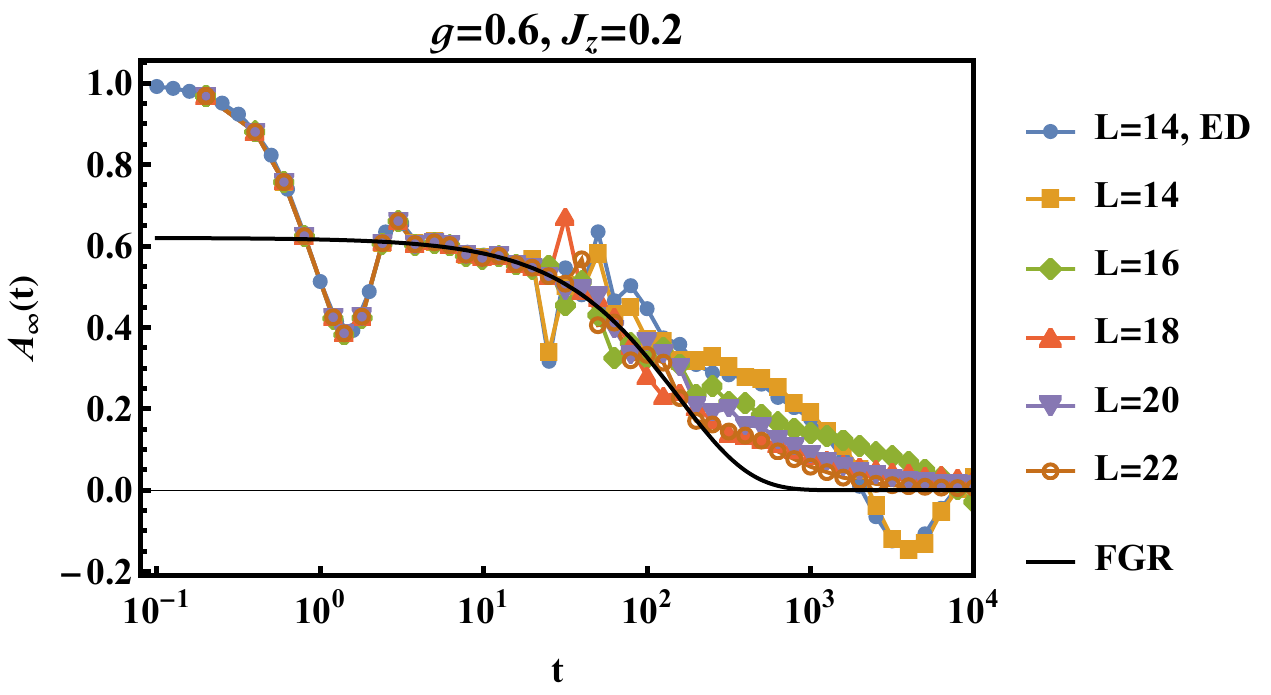} \includegraphics[width=0.45\textwidth]{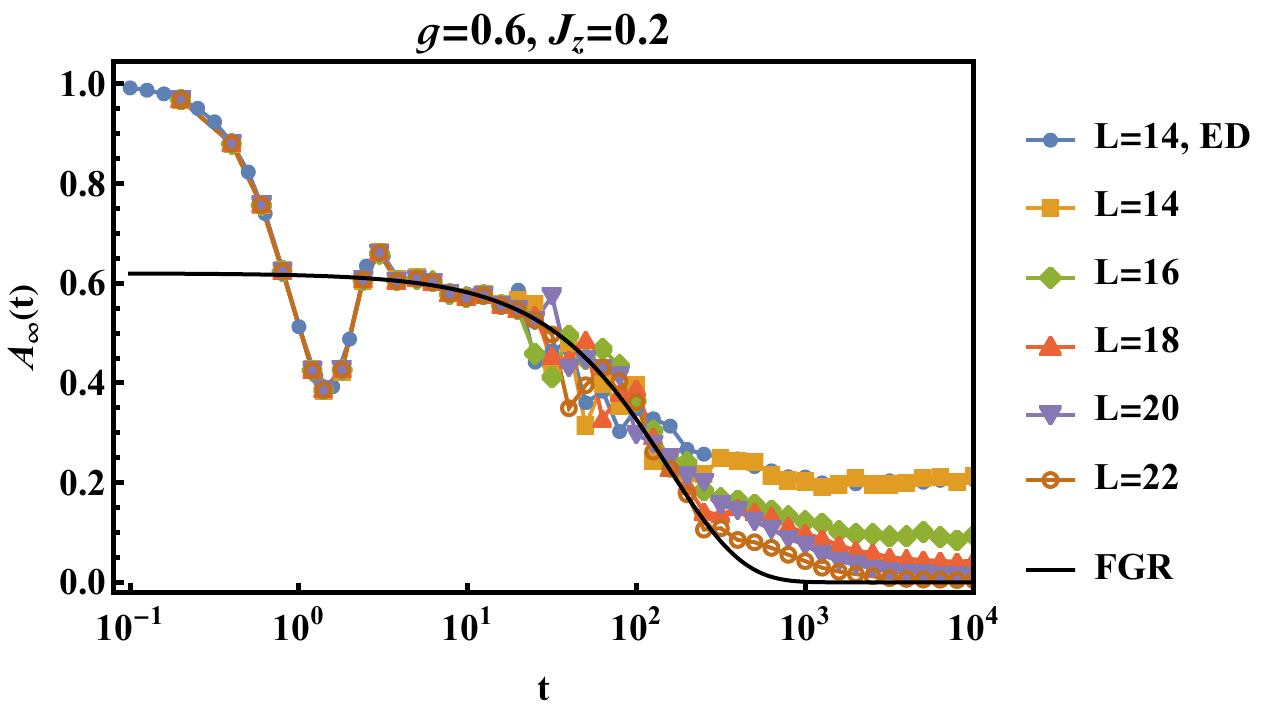}
    \caption{The infinite temperature autocorrelation function for the boundary impurity model (top panel) at $g=0.6$ and $J_z = 0.2$ and the corresponding results with zero transverse-field $g$ on the last site (bottom panel). For large transverse fields, the plateau is blurred and becomes a slowly decaying tail in the top panel. The FGR result is computed numerically and found to be $\Gamma = 0.16J_z^2$. It matches the decay of the edge zero mode (solid black line) after the initial transient but fails when the quasi-conserved operator comes in at late times. Note that the effect of the latter becomes smaller with increasing system size.}
    \label{fig: Autocorrelation, conserved case and FGR}
\end{figure}

We now explain why FGR is valid in the regime of $g=0.6$.  Notice that the transverse-field strength $g$ controls the bandwidth of the bulk quasi-particle spectrum, $\epsilon_k \in [1-g, 1+g]$. The boundary impurity can be written as a four Majorana interaction, $J_z\sigma_1^z\sigma_2^z = -J_za_{1}a_{2}a_{3}a_{4}$. The limiting case for the resonance condition in second-order perturbation, and therefore for FGR to hold, requires an energy-conserving process  where an edge zero mode and one quasi-particle at the top of the band are annihilated and two quasi-particles at the bottom of the band are created,  $2(1-g) = 1+g$. From this argument, the second-order perturbation cannot match the resonance condition for $g < 1/3$. Thus $g=0.6$ clearly places us in a regime where second-order perturbation theory  is valid. In Appendix \ref{Appendix:FGR}, we show that the FGR decay rate is 
\begin{align}
    \Gamma = \frac{1}{2^L} \int_0^{\infty} dt\ \text{Tr}[\Dot{\psi}_0(t)\Dot{\psi}_0(0)],
    \label{Eq: Fermi's Golden Rule 0Mode}
\end{align}
where $\psi_0$ is the zero mode of the TFIM \eqref{Eq: Zero mode of Transverse-field Ising}, we define $\Dot{\psi}_0 = i[J_z\sigma_1^z\sigma_2^z,\psi_0]$ and $\Dot{\psi}_0(t)$ evolves with the unperturbed Hamiltonian $H_0$. 

Fig.~\ref{fig: Autocorrelation, conserved case and FGR} shows the autocorrelation of the boundary impurity (top panel) and the boundary impurity with zero transverse field on the last site (bottom panel). An exponential in time behavior with the FGR decay rate is plotted in both panels, where  
 we numerically compute the FGR decay rate to be $\Gamma = 0.16J_z^2$ for $g=0.6$. This decay rate captures the decay of the autocorrelation function after the initial transient. At late times, the decay of the autocorrelation slows down due to the presence of the quasi-conserved operator. Since we now study a larger transverse-field $g=0.6$, the presence of plateau is not as clear as in Fig.~\ref{fig: autocorrelation g=0.5}. Nevertheless, the bottom panel of Fig.~\ref{fig: Autocorrelation, conserved case and FGR} shows that as $L$ increases, the conserved operator becomes more and more delocalized, and FGR depicts the full decay in the thermodynamic limit.

\begin{figure}[h!]
    \centering \includegraphics[width=0.45\textwidth]{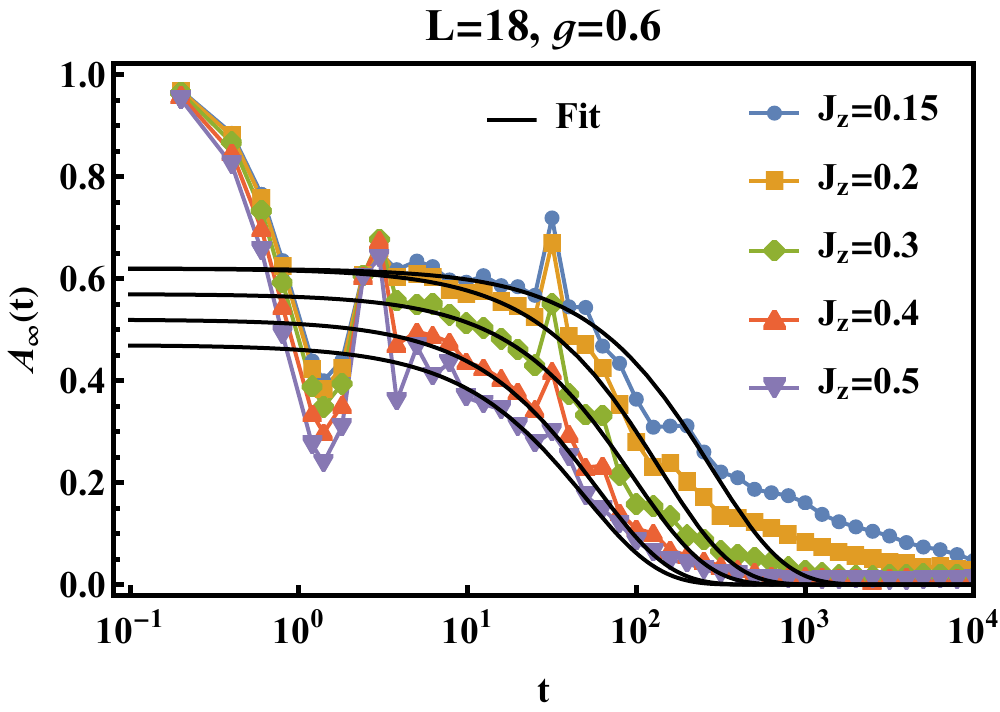} \includegraphics[width=0.45\textwidth]{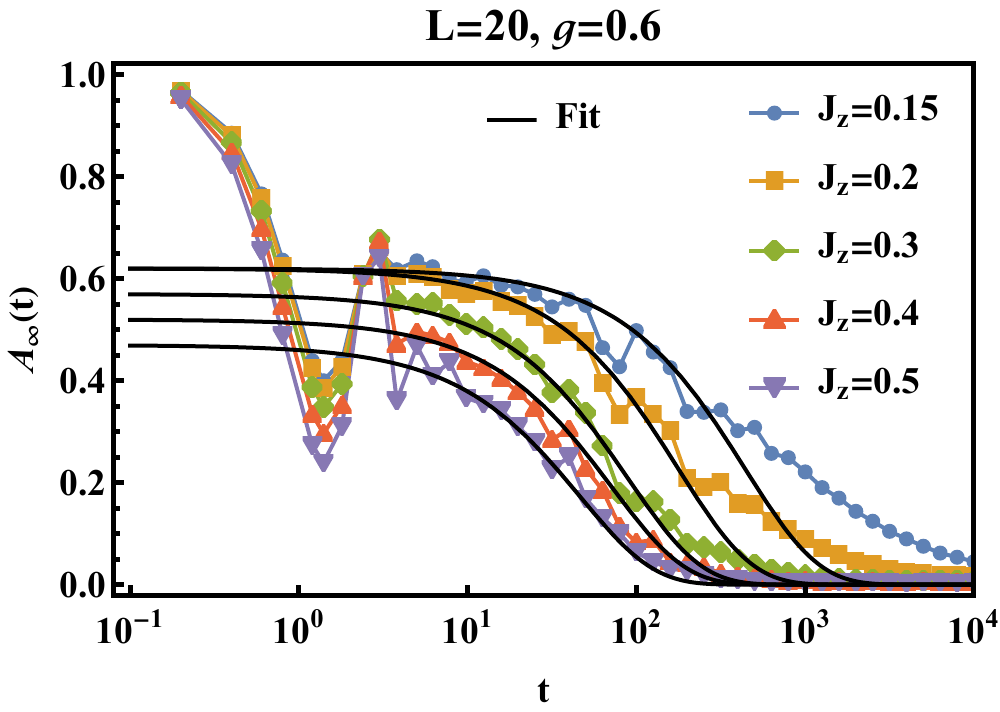} \includegraphics[width=0.45\textwidth]{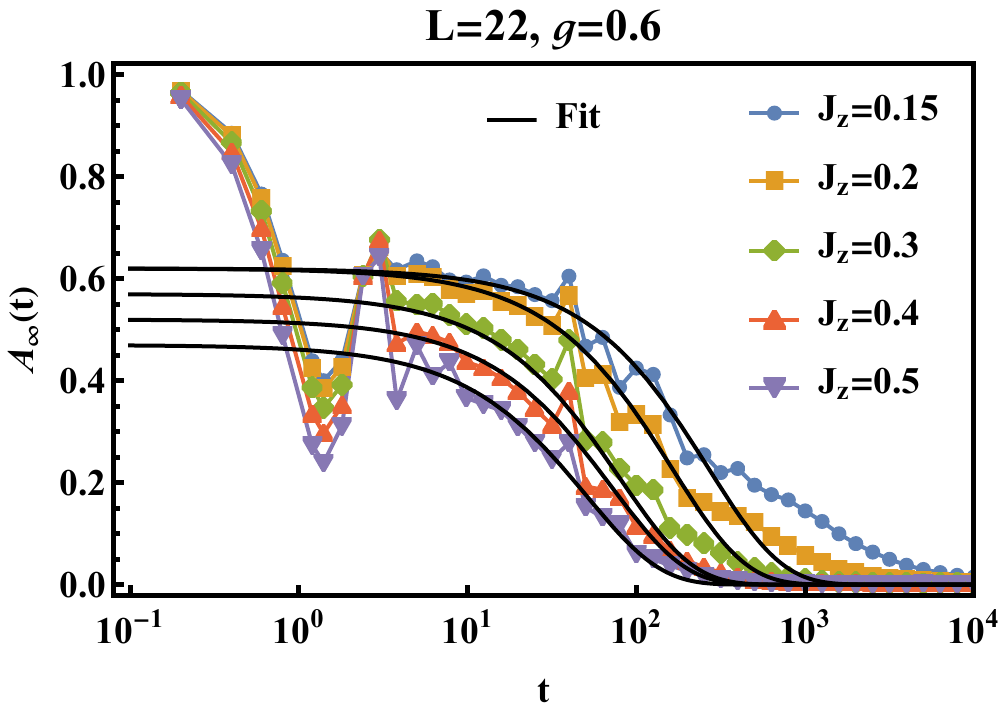}
    \caption{The infinite temperature autocorrelation function for $L = 18$ (top), $20$ (middle) and $22$ (bottom) with $g = 0.6$ and different strengths of the integrability breaking term $J_z$. As $J_z$ decreases, the lifetime increases. Each data set is fitted (solid black line) with an exponential function $C \exp(-t\Gamma_{\text{fit}})$ where the decay rate is determined from the average $\Gamma_{\text{fit}} = (\Gamma_{90\%} + \Gamma_{50\%})/2$, where $\Gamma_{x\%}$ is the inverse time at which the autocorrelation is $0.0x C$.  The decay rate is in units of $J_x=1$.} 
    \label{fig: Autocorrelation in L 18, 20 22}
\end{figure}

\begin{figure}[h]
    \centering \includegraphics[width=0.45\textwidth]{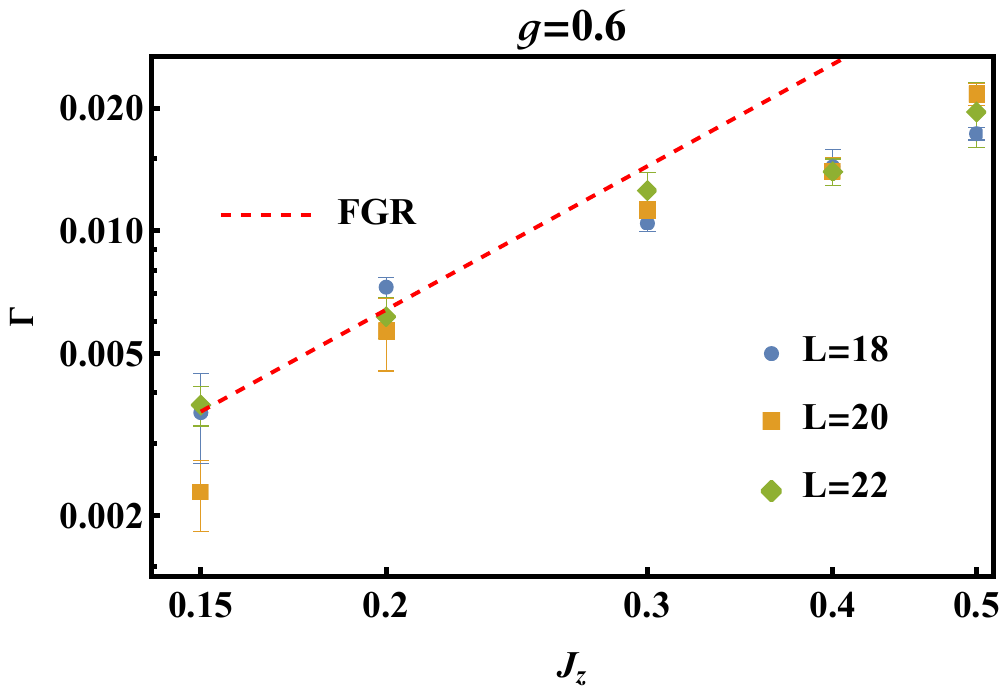} 
    \caption{$\Gamma_{\text{Fit}}$ vs. $J_z$ on a  log-log scale for $L=18, 20, 22$. The decay rate from FRG gives  $\Gamma_{\text{Fit}} = 0.16J_z^2$ for small $J_z$ (dashed red line). The system size effect becomes larger for small $J_z$ as the autocorrelation is strongly influenced by the quasi-conserved operator in Fig.~\ref{fig: Autocorrelation in L 18, 20 22}. The decay rate is in units of $J_x=1$.}
    \label{fig: Decay Rate}
\end{figure}

We explore the autocorrelation function with $J_z$ between $0.15-0.5$ and for system sizes $L=18, 20,22$. The results are summarized in Fig.~\ref{fig: Autocorrelation in L 18, 20 22}, with a numerical fitting to an exponential decay also shown. The fitted decay rate is compared with FGR results in Fig.~\ref{fig: Decay Rate}. For small $J_z$, the decay rate follows the prediction of FGR. The increase of error bars comes from the enhancement of oscillations in the autocorrelation and the late time slowing down of the decay as $J_z$ decreases. This is a finite system size effect. In particular, as $J_z$ decreases, the quasi-conserved operator becomes more localized and more similar to the zero mode $\psi_0$. Therefore, one has to increase the system size further to separate them. 
In appendix \ref{Appendix:Krylov}, we connect the decay of the edge zero mode to a tunneling process of a 1D particle from the edge to the bulk in Krylov space. We show how FGR can be recovered in Krylov space.

In appendix \ref{Appendix:Krylov}, we also highlight a qualitative difference between how the Krylov hopping parameters scale for the impurity model and the more standard non-integrable model with non-zero $J_z$ everywhere. In particular, although both models are non-integrable, the Krylov parameters scale as a square-root for the impurity model, while they scale linearly for the model where $J_z$ is non-zero everywhere.

\section{Conclusions} \label{sec:Conc}

The TFIM is one of the most basic models that hosts an edge zero mode. Understanding the stability of the zero mode to perturbations is important both for practical realizations, as well as for a fundamental understanding of nonequilibrium dynamics. This paper has studied the effect of a weak boundary integrability-breaking perturbation. We have compared this perturbation to the conventional one, where integrability-breaking perturbations are uniformly included all along the chain. We showed a qualitatively different behavior in the dynamics of the zero modes for these two cases.

In particular, for the impurity model, the zero mode decays much more slowly than for the case where the perturbations are non-zero all along the chain. The slow decay arises because the zero mode has an overlap with a quasi-conserved quantity. We explicitly identified this quasi-conserved quantity by a trick that involves local modifications of couplings at the end of the spin chain to enforce the exact degeneracy of the spectrum for any finite system size. We showed that in the thermodynamic limit, the overlap between the zero mode with the quasi-conserved quantity becomes smaller, approaching zero as $L\rightarrow \infty$. In addition, we showed that for large enough transverse fields and in the thermodynamic limit, the zero mode decay could be captured by FGR. 

While we have a quantitative understanding of the zero mode decay for $g\geq 1/3$, an important open question is the fate of the zero mode for small $g$. We do not expect FGR to hold below $g<1/3$. How the decay rate changes as $g$ becomes smaller is left for future studies. The analytic construction of the quasi-conserved quantities presented in Appendix \ref{Appendix:Quasi-conserved operator spin chain} might help these studies. In addition, the Krylov method, generalized to systems in the thermodynamic limit, employed here to recover FGR (see Appendix \ref{Appendix:Krylov}), may also be helpful. Last, it is worth mentioning that the impurity model can be simulated in a noisy intermediate scale quantum device, as was done for the kicked Ising model with open \cite{Abanin-Aleiner22,Harle2023} and duality twisted boundary conditions \cite{samanta2023isolated}.

{\sl Acknowledgments:} 
This work was supported by the US Department of Energy, Office of
Science, Basic Energy Sciences, under Award No.~DE-SC0010821 (HY and AM), by the National Science Foundation under Grant NSF DMR-2116767 (LK and AGA), and by NSF-BSF grant 2020765 (AGA). HY acknowledges the support of the NYU IT High-Performance Computing resources, services, and staff expertise.

\appendix
\section{Operator growth in Krylov space}
\label{Appendix:Krylov}
Besides the direct study of the autocorrelation function to determine the decay rate of a given operator $O_1$, there is another approach for extracting decay rates. This involves studying how the operator evolves and spreads in operator space. First, the Heisenberg time evolution of the operator $O_1$ under the Hamiltonian $H$ is
\begin{align}
    O_1(t) = e^{iHt}O_1 e^{-iHt} = \sum_{n=0}^{\infty} \frac{(it)^n}{n!} \mathcal{L}^n O_1,
\end{align}
where we define $\mathcal{L}O = [H,O]$ for any operator $O$. In operator space, we treat the operator $O_1$ as a vector $|O_1)$, and $\mathcal{L}$ is called the superoperator since it is an operator which acts on operators. In this new notation, the time evolution of $O_1$ becomes
\begin{align}
    |O_1(t)) = e^{i\mathcal{L}t}|O_1),
    \label{Eq: time evolution of operator}
\end{align}
where $\mathcal{L}$ plays the role of a ``Hamiltonian'' as it is the generator of time evolution for the operators. The operator space is spanned by the set of operators generated by $\mathcal{L}$ acting on $|O_1)$: $\{ |O_1), \mathcal{L}|O_1), \mathcal{L}^2|O_1), \ldots\}$, and is called the Krylov space. The inner product between two operators $A$ and $B$ is defined as 
\begin{align}
    (A|B) = \frac{1}{2^L} \text{Tr}[A^\dagger B].
\end{align}

To construct an orthonormal basis, we apply the Lanczos algorithm. Starting from a normalized operator $|O_1)$, one can generate a new basis element $|O_2)$ via  $\mathcal{L}|O_1) = b_1|O_2)$ with $b_1 = \sqrt{|\mathcal{L}|O_1)|^2}$, the norm of $\mathcal{L}|O_1)$. The remaining basis elements are computed from the iterative relation for $n \geq 2$
\begin{align}
    \mathcal{L}|O_n) = b_n|O_{n+1}) + b_{n-1}|O_{n-1}),
\end{align}
where $b_n = \sqrt{|\mathcal{L}|O_n)-b_{n-1}|O_{n-1})|^2}$.
Finally, one can represent $\mathcal{L}$ as a tri-diagonal matrix in this basis
\begin{align}
    \mathcal{L} =
    \begin{pmatrix}
        0 & b_1 \\
        b_1 & 0& b_2 \\
        & b_2 & 0 & \ddots\\    &  & \ddots & \ddots 
    \end{pmatrix}.
    \label{Eq: tridiagonal Krylov}
\end{align}
In the following, we refer to this tridiagonal matrix as the Krylov Hamiltonian.

There are two kinds of representations in the numerical computation of the off-diagonal elements $\{b_n\}$: matrix representation or Pauli strings. In the matrix representation, $|O_1)$ is a $2^L \times 2^L$ matrix. It is usually sparse if $|O_1)$ is some local operator, e.g., $\sigma_1^x$. After some iterations, one begins to generate non-sparse matrices $|O_n)$, and the computation is limited by the memory to store such matrices. The non-sparsity of the basis $|O_n)$ is a property both for integrable and non-integrable models unless, for the former, a suitable Majorana basis is available to perform the expansion.

The idea behind the Pauli strings representation is to overcome the non-sparsity of the matrix representation, and below we summarize the discussion in \cite{parker2019universal}. For spin systems, a Pauli string is a series of tensor products of Pauli matrices on each site as follows
\begin{align}
    i^\delta (-1)^\epsilon (\sigma_1^z)^{v_1}(\sigma_1^x)^{w_1} \otimes \ldots \otimes (\sigma_L^z)^{v_L}(\sigma_L^x)^{w_L},
\end{align}
where $\delta, \epsilon, v_n, w_n \in \{0, 1\}$. Thus one only requires to store $2L+2$ numbers and each of them is either 0 or 1, for a given Pauli string. Since $\sigma^x\sigma^z = -i\sigma^y $ and the identity corresponds to setting $v = w = 0$, the Pauli string representation indeed exhausts all possible combinations of local spin operators. For two given Pauli strings $\sigma$ and $\sigma'$, labeled by $\{ \delta, \epsilon, \vec{v}, \vec{w} \}$ and $\{ \delta', \epsilon', \vec{v'}, \vec{w'} \}$, the new Pauli string generated from the commutation $\sigma'' = [\sigma,\sigma']$ obeys the algebra rules
\begin{align}
    &\delta'' = \delta + \delta'\ \text{mod}\ 2,\\
    &\epsilon'' = \epsilon +\epsilon' + \delta\delta' + \Vec{w}\cdot \Vec{v'}\ \text{mod}\ 2,\\
    &\Vec{v''} = \Vec{v} + \Vec{v'}\ \text{mod}\ 2,\\
    &\Vec{w''} = \Vec{w} + \Vec{w'}\ \text{mod}\ 2.
\end{align}
For an operator represented by Pauli strings, there are overall $(2L+2)\times N$ numbers, where $N$ is the number of Pauli strings since it requires another $N$-dimensional vector to store the coefficients of each Pauli string. As long as $N$ is much smaller than  $2^L \times 2^L$, the Pauli string representation is efficient in the memory cost. However, it is time-consuming to add or subtract operators which consist of many Pauli strings because one has to scan through all the Pauli strings of each operator to determine if the two operators contain the same Pauli string. Addition and subtraction are much simpler operations in matrix representation. Therefore, one may choose either the matrix or the Pauli string representation in numerical computation depending on how fast the number of Pauli strings grows and how many off-diagonal elements $b_n$ one needs to compute.

\begin{figure}[h]
    \centering \includegraphics[width=0.45\textwidth]{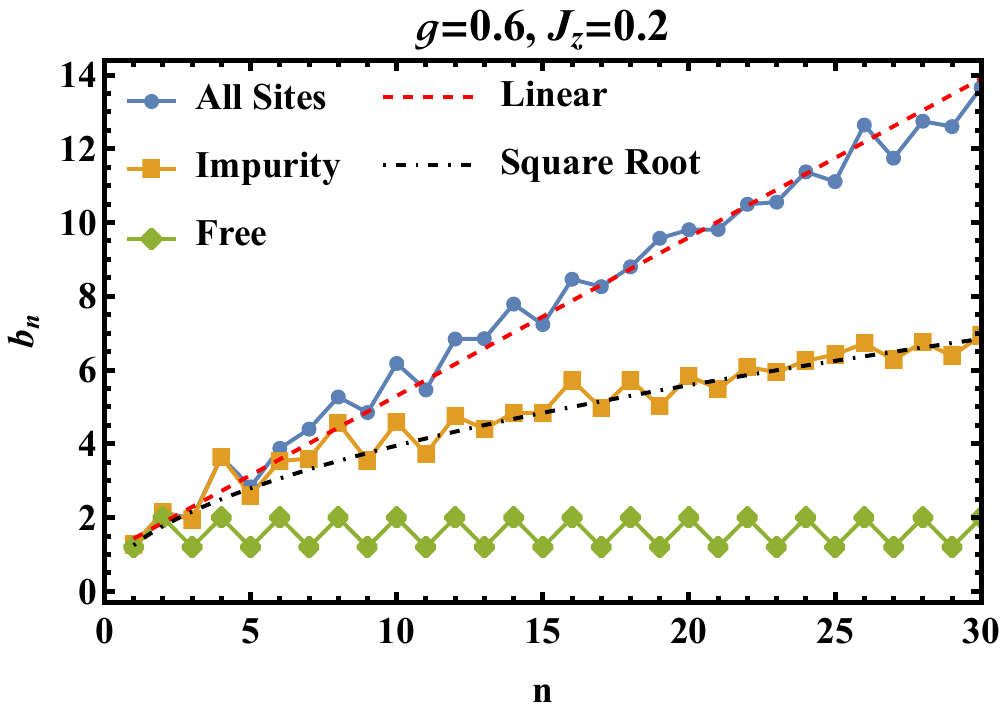}
    \caption{Off diagonal matrix element $b_n$ of the Krylov Hamiltonian for the seed operator $\sigma^z_1$ and for the transverse-field Ising model with different perturbations. Without perturbations, $b_n$ is dimerized. For the boundary impurity $J_z\sigma_1^z\sigma_2^z$, $b_n$s follow a square root growth.  With perturbation on all sites $J_z\sum_{i}\sigma_i^z\sigma_{i+1}^z$, $b_n$s grow linearly.}
    \label{fig: Krylov}
\end{figure}

In Fig.~\ref{fig: Krylov}, we show the system size independent results of $\{b_n\}$ generated by $|O_1) = \sigma_1^x$. Numerically, we calculate  $\{b_n\}$ with increasing system size until $\{b_n\}$ is independent of system size.
We employ the matrix representation for the model with perturbations on all sites, and we employ the Pauli strings representation for the boundary impurity and the free case. The growth of $\{b_n\}$ reflects the integrability of the system. Without any perturbation, the system is free and $\{b_n\}$ are perfectly dimerized, which allows for an exactly conserved zero mode localized on the first site. In 
 particular, \eqref{Eq: tridiagonal Krylov} becomes a Su–Schrieffer–Heeger (SSH) model with topologically non-trivial dimerization. However, the perfect dimerization is altered by interactions and it is argued \cite{parker2019universal} that a linear growth appears when the system is chaotic, e.g. perturbations on all sites of a chain in Fig.~\ref{fig: Krylov}. When integrability is broken at the boundary, we observe a square root behavior of $\{b_n\}$. The square root behavior is also seen in the integrable interacting model of the XXX chain, see \cite{parker2019universal}. In what follows, as suggested by the numerics in Fig.~\ref{fig: Krylov}, we assume a square root growth of the $b_n$ for the boundary impurity model superimposed on a non-zero dimerization, and we use this property for building a toy model.

\begin{figure*}[t]
    \centering \includegraphics[width=0.32\textwidth]{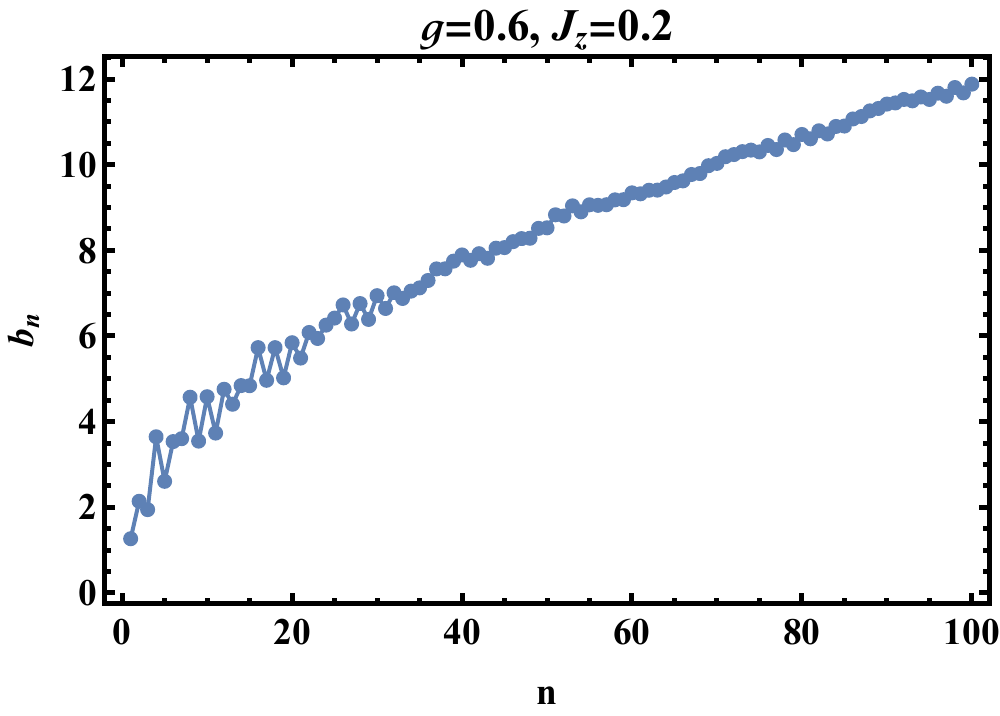} \includegraphics[width=0.32\textwidth]{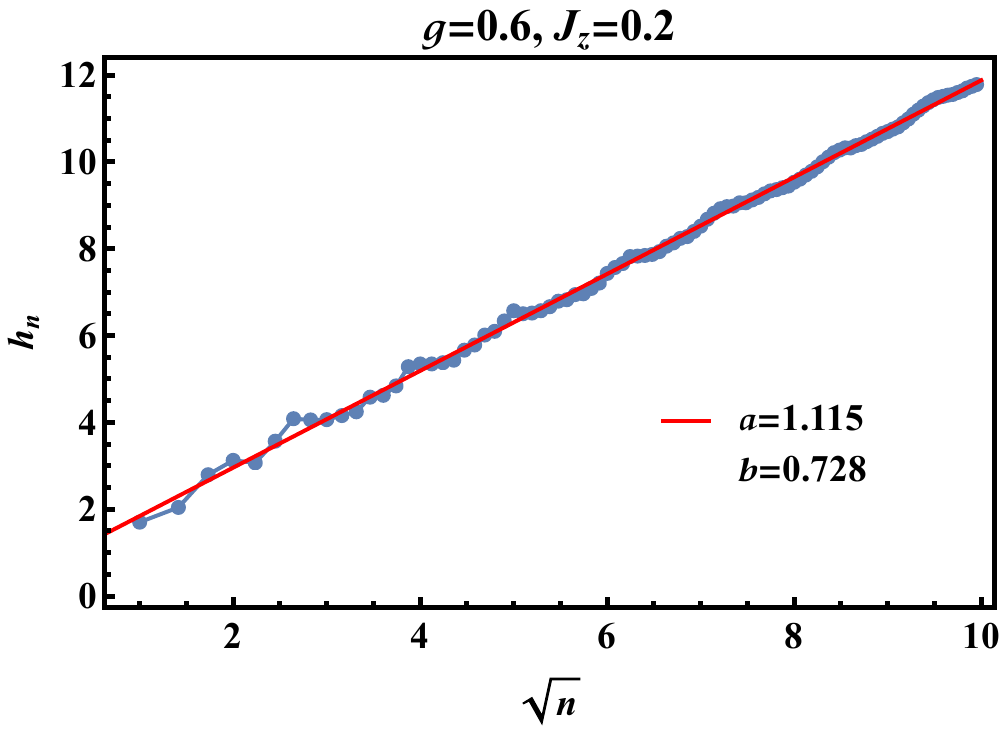} \includegraphics[width=0.32\textwidth]{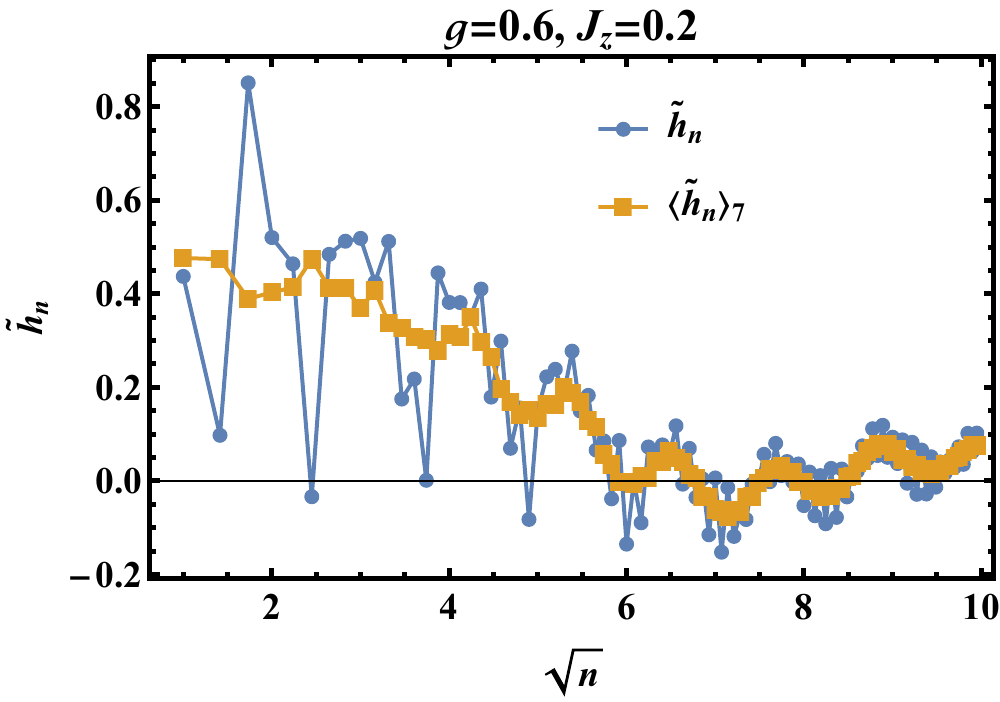}
    \caption{Off-diagonal matrix element $b_n$ (left panel). $h_n$ (middle panel) and $\Tilde{h}_n$ (right panel) are generated according to \eqref{Eq: hn} and \eqref{Eq: htn}. As $b_n$s follow square root growth, both $h_n$ and $\Tilde{h}_n$ are plotted on the $\sqrt{n}$-scale. $h_n$ describes the average growth of $b_n$ and is fitted with $a\sqrt{n}+b$. $\Tilde{h}_n$ illustrates the dimerization of $b_n$, which only survives up to $\sqrt{n} \sim 6$. Also plotted is the moving 7-sites average $\langle \Tilde{h}_n \rangle_7$ as a smooth approximation for the $\Tilde{h}_n$.}
    \label{fig: h ht dimerization}
\end{figure*}

To understand how $\{b_n\}$ is related to the decay rate of edge zero mode, we follow and summarize the discussion in  Refs.~\onlinecite{Yates20,Yates20a}. First, one can write the Schr{\"o}dinger equation of the operator from \eqref{Eq: time evolution of operator} and \eqref{Eq: tridiagonal Krylov}
\begin{align}
    -i\partial_t \Psi_n = b_n \Psi_{n+1} + b_{n-1}\Psi_{n-1},
    \label{Eq: SSH-like equation}
\end{align}
where $\Psi_n$ are the coefficients of the operator $\Psi$ expanded in Krylov space, $|\Psi) = \sum_{n=1}^\infty \Psi_n|O_n)$. 
Since for the non-interacting case, the $b_n$ are perfectly dimerized, we proceed by
decomposing $\Psi_n$ and $b_n$ into two parts
\begin{align}
     &\Psi_n = i^n \left[\alpha_n + (-1)^n \Tilde{\alpha}_n \right],\\
     &b_n = h_n + (-1)^n\Tilde{h}_n,
\end{align}
where $h_n$ depicts the average growth of $b_n$ and $\Tilde{h}_n$ senses the dimerization of $b_n$. The Schr{\"o}dinger equation then becomes
\begin{align}
    -&i\partial_t \left[\alpha_n + (-1)^n\Tilde{\alpha}_n \right] \\
    =& i\left[ (h_n\alpha_{n+1} - \Tilde{h}_n\Tilde{\alpha}_{n+1} - h_{n-1}\alpha_{n-1} - \Tilde{h}_{n-1}\Tilde{\alpha}_{n-1} )\nonumber \right. \\
     &\left. +(-1)^n(\Tilde{h}_n\alpha_{n+1} -h_n\Tilde{\alpha}_{n+1} + h_{n-1}\Tilde{\alpha}_{n-1} + \Tilde{h}_{n-1}\alpha_{n-1})  \right].
     \nonumber
\end{align}
Since $(-1)^n$ is rapidly oscillating, the Schr{\"o}dinger equation can be solved by equating the terms with and without $(-1)^n$ on both sides. Now, we assume $\alpha_n, \Tilde{\alpha}_n, h_n$ and $\Tilde{h}_n$ are slowly varying and smooth functions of $n$. In the continuous limit of $n$, we expan $h_{n\pm 1} \approx h(n) \pm \partial_n h(n)$ and the same expansions for $\Tilde{h}_{n\pm 1}, \alpha_{n\pm 1}$ and $\Tilde{\alpha}_{n\pm 1}$. In addition,  we only keep terms up to one derivative. One obtains
\begin{align}
    -i\partial_t
    \begin{pmatrix}
        \alpha\\
        \Tilde{\alpha}
    \end{pmatrix}
    = i
    \begin{pmatrix}
      \partial_n h + 2h\partial_n & -2\Tilde{h}+\partial_n \Tilde{h}\\
      2\Tilde{h}-\partial_n \Tilde{h} & -\partial_n h - 2h\partial_n
    \end{pmatrix}
    \begin{pmatrix}
        \alpha\\
        \Tilde{\alpha}
    \end{pmatrix}.
\end{align}
The diagonal terms can be massaged into simple linear spatial derivatives. First by rescaling fields, $(\alpha\ \Tilde{\alpha})^T = \chi/\sqrt{h}$, $\partial_n h$ is canceled. Then, absorbing $2h$ into $n$ via the change of variables
\begin{align}
    X = \int_0^n \frac{d n'}{2h(n')}.
    \label{Eq: spatial coordinates X}
\end{align}
one finally arrives at
\begin{align}
    -i\partial_t \chi = \left[ -i\sigma_z \partial_X + \sigma_y m(X) \right] \chi,
    \label{Eq: Dirac Toy model}
\end{align}
where $m(X) = 2\Tilde{h} - (\partial_X \Tilde{h})/2h$. Essentially, we have approximated the generalized SSH model in Krylov space as a continuous 1D Dirac equation with spatially non-uniform mass that contains information about the dimerization of $b_n$.  In the following, we first extract the information from numerical results and then apply the above toy model to compute how the decay rate is influenced by the boundary impurity.

Fig.~\ref{fig: h ht dimerization} shows the system size independent $b_n$ up to $n=100$ with $g=0.6$ and $J_z=0.2$, and determines $h_n$ and $\Tilde{h}_n$ from the $b_n$ as follows
\begin{align}
    &h_n \approx \frac{b_n + b_{n+1}}{2}, \label{Eq: hn} \\
    &\Tilde{h}_n \approx (-1)^n \frac{b_n - b_{n+1}}{2}.
    \label{Eq: htn}
\end{align}
Here we present $h_n$ and $\Tilde{h}_n$ in $\sqrt{n}$ scale since the $b_n$ follow a square root growth. One can approximate $h_n$ (middle panel) as $h_n \approx a\sqrt{n}+b$ and the new spatial coordinate $X$ from \eqref{Eq: spatial coordinates X} is
\begin{align}
    X = \frac{\sqrt{n}}{a} - \frac{b \ln (1+a\sqrt{n}/b)}{a^2}.
\end{align}

\begin{figure}[h]
    \centering \includegraphics[width=0.45\textwidth]{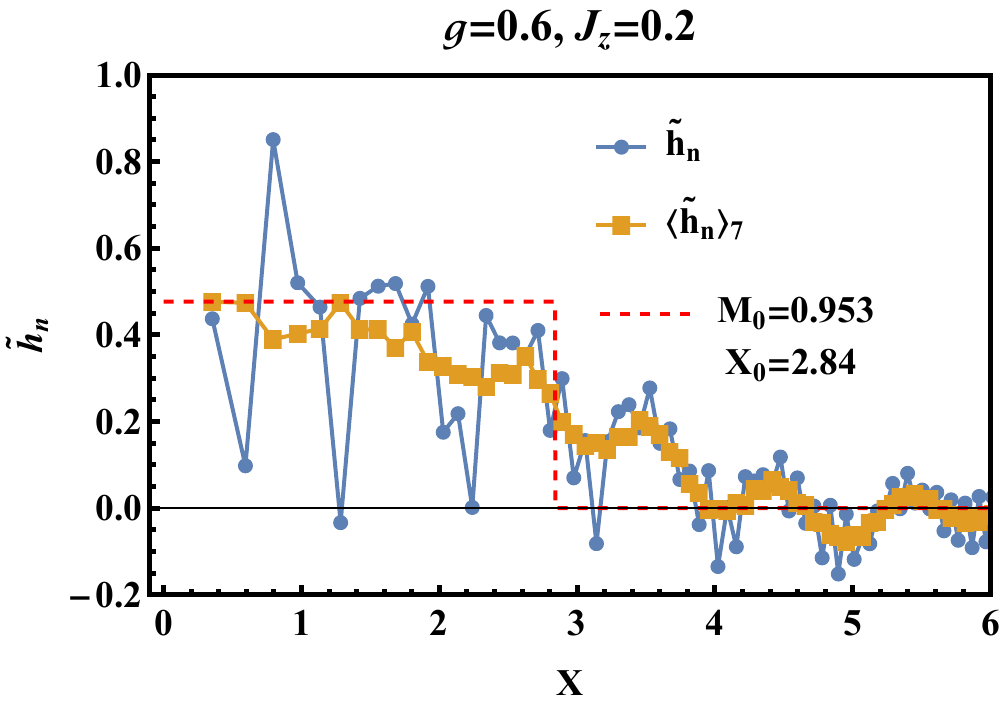}    \caption{$\Tilde{h}_n$ vs coordinate $X$. The 7-sites moving average $\langle \Tilde{h}_n \rangle_7$ is fitted with the step function $M_0 \theta (X_0 -X)/2$, where $M_0/2$ is fitted from the value of the first $\langle \Tilde{h}_n \rangle_7$ or the maximum of $\langle \Tilde{h}_n \rangle_7$. In this case, they happen to be the same, giving one fitting result. $X_0$ is fitted by extrapolating the $\langle \Tilde{h}_n \rangle_7$ data to find the smallest $X_0$ where $\langle \Tilde{h}_n \rangle_7$ becomes $M_0/4$.}
    \label{fig: ht in coordinate X}
\end{figure}

For $\Tilde{h}_n$ and its $7$ site moving average $\langle \Tilde{h}_n \rangle_7$ (right panel), the dimerization only survives up to $\sqrt{n} \sim 6$, and therefore the edge zero mode has to decay eventually. The moving average $\langle \Tilde{h}_n \rangle_7$ mimics the slowly varying continuous $h(n)$ in the toy model. In the new coordinate $X$, $\Tilde{h}_n$ and $\langle \Tilde{h}_n \rangle_7$ is presented in Fig.~\ref{fig: ht in coordinate X}. To determine the mass of the toy model and perform analytic calculations, we first approximate the mass by $m(X) \approx 2\Tilde{h}(X)$. This is because the moving average $\langle \Tilde{h}_n \rangle_7$ is rather smooth and $h$ grows with $X$, so that $(\partial_X \Tilde{h})/2h$ can be dropped. Then, we fit $\langle \Tilde{h}_n \rangle_7$ with a step function profile $ M_0 \theta(X_0 - X)/2$ for analytic simplicity.  Thus the mass is approximated to be $m(X) \approx 2\Tilde{h}(X) \approx M_0 \theta(X_0 - X)$. 

\begin{figure}[h]
    \centering \includegraphics[width=0.45\textwidth]{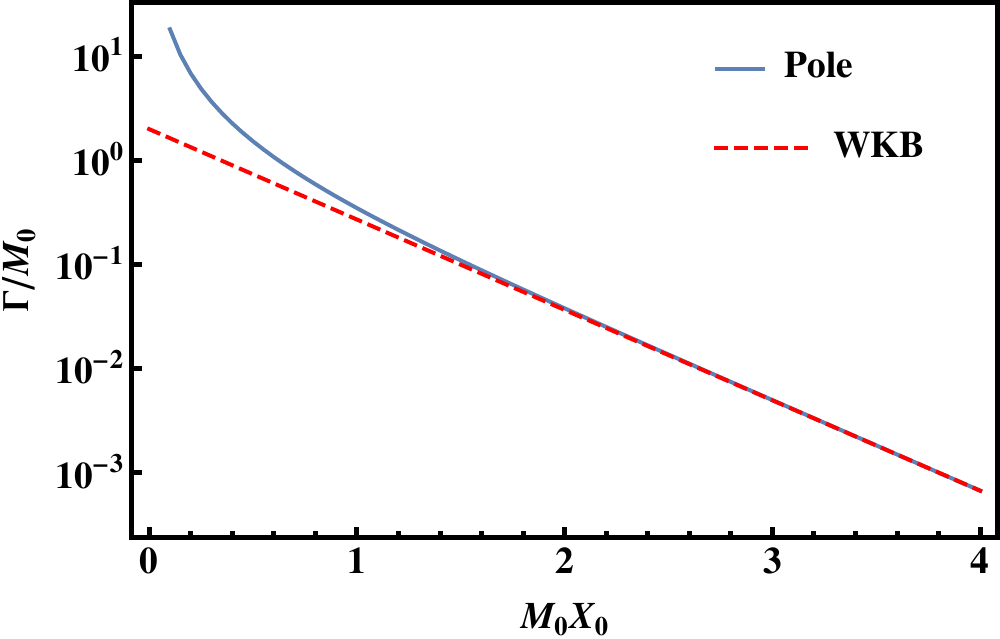}
    \caption{Comparison between decay rates from the pole of the Green's function \eqref{Eq: Pole} and from WKB  \eqref{Eq: WKB}. Although WKB is typically valid when $M_0X_0 \gg 1$, the numerical results already show a good agreement for $M_0 X_0 > 2$.}
    \label{fig: WKBvsPole}
\end{figure}

For a given mass distribution, one can solve the Green's function of the toy model \eqref{Eq: Dirac Toy model},  and from that extract the decay rate of the edge zero mode  from the pole of the Green's function on the positive imaginary axis in the complex plane. To find the pole of the Green's function, one can solve the scattering problem because the transmission and reflection coefficients share the same poles as the Green's function. The scattering solution for an incident wave coming from  $X = \infty$, with a step function mass distribution, for $X < X_0$ is:
\begin{align}
    &\chi (X < X_0) \nonumber\\
    &= e^{iEt} \left[ A_1 e^{-\kappa X} 
    \begin{pmatrix}
        -iM_0\\
        i\kappa + E
    \end{pmatrix}
    + A_2 e^{\kappa X}
    \begin{pmatrix}
        -iM_0\\
        -i\kappa + E
    \end{pmatrix}
    \right],
\end{align}
and for $X > X_0$ is:
\begin{align}
    &\chi (X > X_0) \nonumber\\
    &= e^{iEt} \left[ e^{iE( X-X_0)} 
    \begin{pmatrix}
        0\\
        1
    \end{pmatrix}
    + B e^{-iE( X-X_0)}
    \begin{pmatrix}
        1\\
        0
    \end{pmatrix}
    \right],
\end{align}
where $\kappa = \sqrt{M_0^2-E^2}$. The coefficients, $A_1, A_2$ and $B$, are determined by boundary conditions at $X = 0$ and $X_0$. In the original discrete Schr{\"o}dinger equation \eqref{Eq: SSH-like equation}, the boundary condition at $n = 0$ is $\Psi_0 = \alpha_0 + \Tilde{\alpha}_0 = 0$. In terms of the toy model, one obtains the boundary condition at $X = 0$: $\sigma_x\chi(0) = -\chi(0)$. At $X =X_0$, the wave function is continuous: $\chi(X \rightarrow X_0^-) = \chi(X \rightarrow X_0^+)$. From these two conditions, the coefficients can be solved for and they share the common factor in the denominator. The poles are the value of $E$ at which the common denominator vanishes,
\begin{align}
    \kappa \cosh(\kappa X_0)-m \sinh(\kappa X_0) + iE \sinh(\kappa X_0) = 0.
    \label{Eq: Pole}
\end{align}
For the decay rate of edge zero mode, one is looking for the solution $E = i\Gamma$ of the above equation. In the limit $\Gamma/M_0 \ll 1$ and $M_0 X_0 \gg 1$, the solution is the WKB approximation
\begin{align}
    \Gamma \approx 2M_0 e^{-2M_0 X_0}.
    \label{Eq: WKB}
\end{align}
We compare the numerical results of the poles with the WKB formula in Fig.~\ref{fig: WKBvsPole}, and find that they are in good agreement at $M_0 X_0 > 2$.

Although we have performed a crude approximation to establish the toy model and extract information from $h_n$ and $\Tilde{h}_n$, the underlying physical picture is quite simple. The decay of the edge zero mode can be realized as a tunneling event. Without integrability-breaking perturbations, the system has perfect dimerization and $X_0 \rightarrow \infty$ so that the zero mode has an infinitely long lifetime. With perturbations, the dimerization terminates at some finite $X_0$ and the edge zero mode becomes unstable as it can now tunnel through the finite potential barrier $M_0$. When one gradually turns off the perturbation $J_z$, approaching the free limit, $X_0$ strongly depends on $J_z$ as $X_0 \rightarrow \infty$ with $J_z \rightarrow 0$, but $M_0$ stays around some $\mathcal{O}(1)$ number. Therefore, the $J_z^2$ dependence in the FGR region is expected to arise from the exponential factor of the WKB result, i.e., we expect $M_0 X_0 \propto - \ln J_z$,
where the $J_z$-dependence primarily comes from $X_0$.

\begin{figure*}
    \centering \includegraphics[width=0.96 \textwidth]{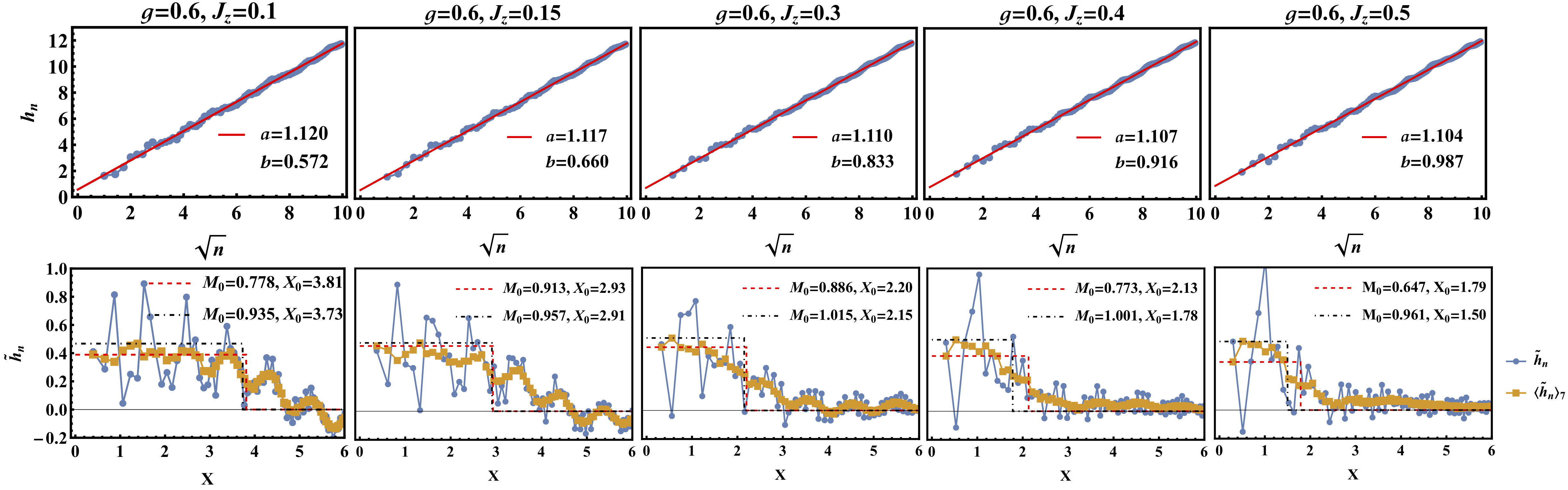}
    \caption{$h_n$ (top panels) and $\Tilde{h}_n$ (bottom panels) for $g = 0.6$ with different boundary impurity strengths $J_z$. $h_n$ is fitted with $a\sqrt{n}+b$. The 7-sites moving average $\langle \Tilde{h}_n \rangle_7$ is fitted with the step function $M_0 \theta (X_0 -X)/2$, where $M_0/2$ can is fitted from the value of the first $\langle \Tilde{h}_n \rangle_7$ or by its maximum value. $X_0$ is fitted by extrapolating the $\langle \Tilde{h}_n \rangle_7$ data to find the smallest $X_0$ where $\langle \Tilde{h}_n \rangle_7$ becomes $M_0/4$.  This leads to two different fitting results. }
    \label{fig: h and ht}
\end{figure*}

\begin{figure}[h]
    \centering \includegraphics[width=0.45\textwidth]{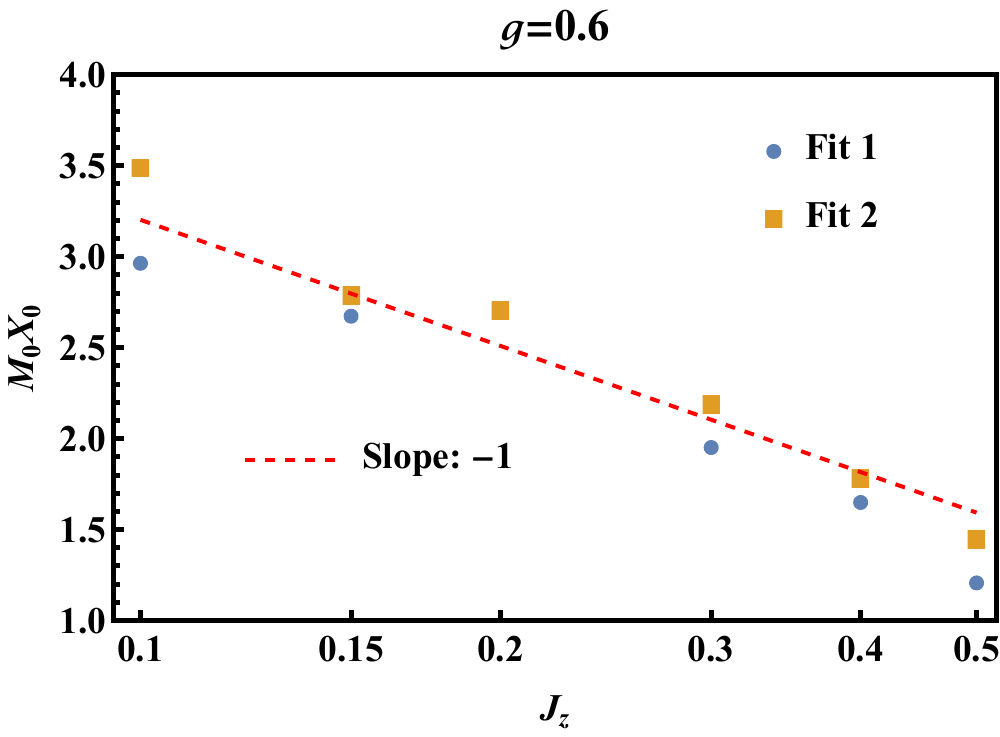}
    \caption{$M_0X_0$ vs $J_z$. The results come from fitting $M_0$ from Fig.~\ref{fig: h and ht} in two different ways. In Fit 1, $M_0$ equals the first $\langle \Tilde{h}_n \rangle_7$ while in Fit 2 $M_0$ equals the maximum of $\langle \Tilde{h}_n \rangle_7$. The plot shows $M_0 X_0 \propto -\ln J_z$, in agreement with a decay rate $\propto J_z^2$.}
    \label{fig: M0X0 vs Jz}
\end{figure}

Fig.~\ref{fig: M0X0 vs Jz} shows results from the two different fittings of Fig.~\ref{fig: h and ht}. It shows the trend $M_0 X_0 \propto -\ln J_z$, supporting the FGR argument in the main text. One advantage of Krylov space is that one can probe the small $J_z$ region more easily than calculating the autocorrelation function. We take $J_z$ to be as small as $J_z = 0.1$ in Figures \ref{fig: M0X0 vs Jz} and \ref{fig: h and ht}, but this regime is rather unfeasible for the autocorrelation function which shows strong system size dependence. However, the decay rate from the toy model is sensitive to the way the fitting is done, as shown in Fig.~\ref{fig: M0X0 vs Jz}. As the oscillations of $\langle \Tilde{h}_n \rangle_7$ become stronger for small $J_z$, fitting $M_0$ and $X_0$ from a step function is rather ambiguous. A more careful analysis of the numerical data is required for small $J_z$. For the region of $g,J_z$ we have explored, we conclude that the decay rate obeys FGR.

\section{Quasi-conserved operator for finite spin chains} \label{Appendix:Quasi-conserved operator spin chain}

The analysis of the edge spin autocorrelation function in section \ref{sec:Quasi-conserved quantity} revealed the existence of a non-local quasi-conserved operator $O_c$ responsible for the observed plateau at intermediate times. Here we give the explicit construction of this operator from the commutation algebra of a finite spin chain. We consider the two Hamiltonians
\begin{align}
    \tilde{H}^A_L&=\sum_{i=1}^{L-1}\sigma^x_i\sigma^x_{i+1}+g\sum_{i=1}^{L-1}\sigma^z_i+J_z\sigma^z_1\sigma^z_2,\label{eq:himp}\\
    \tilde{H}^B_L&=\sum_{i=1}^{L-1}\sigma^x_i\sigma^x_{i+1}+g\sum_{i=1}^{L-1}\sigma^z_i+J_z\sum_{i=1}^{L-2}\sigma^z_i\sigma^z_{i+1}.\label{eq:hint}
\end{align}
Above, the first Hamiltonian \eqref{eq:himp} is the TFIM perturbed by boundary interactions, and with $g=0$ on the last site. The second
Hamiltonian \eqref{eq:hint} has interactions on all sites, except the last site where both $g = J_z = 0$. It is clear that both Hamiltonians commute with the $\mathbb{Z}_2$ symmetry (parity) operator $\mathcal{D}=\sigma^z_1\sigma^z_2...\sigma^z_L$.  

One can attempt to find a conserved operator $O_c$ satisfying
\begin{align}
    &[\tilde{H}^{A,B}_L,O^{A,B}_c]=0, &\{\mathcal{D},O^{A,B}_c\}=0.\label{eq:hcommute}
\end{align}
One way to proceed is to expand $O_c$ as a power series
\begin{equation}
    O_c=\sum_{n=0}^{\infty}J_z^n\left(\sum_{m=0}^{\infty}g^m O_{c}^{(n,m)}\right),\label{eq:seriesJg}
\end{equation}
starting from $O_{c}^{(0,0)}=\sigma^x_1$, similar to the approach in \cite{FendleyXYZ,Else16}. As shown in these references, equation (\ref{eq:hcommute}) can be solved at any order in $g, J_z$, though the series has to be truncated due to the rapid growth in the number of terms.

In the following, we outline another procedure for solving for $O_c$, which is mathematically equivalent to the time average construction in Sec.~\ref{sec:Quasi-conserved quantity}. For a given system size $L$, $O_c$ is a $2^L\times2^L$ matrix and can be expressed as a linear combination of $2^L\times2^L$ orthonormal Pauli string operators,

\begin{align}
    O_c=\alpha_1\sigma^x_1+\alpha_2\sigma^y_1+\alpha_3\sigma^z_1\sigma^x_2+\alpha_4\sigma^z_1\sigma^y_2+\ldots
    \label{Eq: Oc in terms of Pauli Strings}
\end{align}
To determine the coefficients $\{\alpha\}$, one has to solve a homogeneous equation, \begin{align}
    \mathcal{L}_{\Tilde{H}}|O_c) = 0,
    \label{Eq: Oc homogeneous eq}
\end{align}
where $\mathcal{L}_{\Tilde{H}}|O_c) = [\Tilde{H},O_c]$ and $|O_c)$ is a vector with $2^L\times 2^L$ coefficients $\{\alpha\}$. By setting $\Tilde{H} = \Tilde{H}_L^{A}$ or $\Tilde{H} = \Tilde{H}_L^{B}$, one can solve for the corresponding $O_c$ in these two models. The solution of $|O_c)$ is the linear combination of eigenvectors in the zero eigenvalue sector of $\mathcal{L}_{\Tilde{H}}$. In Sec.~\ref{sec:Quasi-conserved quantity}, $O_c$ is defined as the time average of $\sigma_1^x(t)$. In the vector representation of the operator, $|\sigma_1^x(t)) = \exp{[i\mathcal{L}_{\Tilde{H}}t]}|\sigma_1^x)$. Therefore, the time average can be realized as the projection onto the zero eigenvalue sector of $\mathcal{L}_{\Tilde{H}}$,
\begin{align}
    O_c \propto \lim_{T\rightarrow\infty}\frac{1}{T}\int_0^T dt e^{i\mathcal{L}_{\Tilde{H}}t} |\sigma_1^x) = \sum_{j=1}^{N} |\lambda_j)(\lambda_j|\sigma_1^x),
    \label{Eq: Oc projected to zero sector}
\end{align}
where $\{|\lambda_j)\}$ are orthonormal eigenvectors of $\mathcal{L}_{\Tilde{H}}$ with zero eigenvalue and $N$ is the size of this sector. The terms with $|\sigma_1^x)$ projected onto non-zero eigenvalue sectors oscillate in time and will vanish in the time average. According to \eqref{Eq: Oc projected to zero sector}, one obtains another equivalent construction of $O_c$. 

Brute force diagonalization of $\mathcal{L}_{\Tilde{H}}$ is much less efficient than time averaging for large system sizes. Although \eqref{Eq: Oc homogeneous eq} can be reduced to a smaller vector space by the constraint of anticommutation with $\mathbb{Z}_2$ parity, e.g., a single $\sigma_1^z$ operator actually does not contribute in \eqref{Eq: Oc in terms of Pauli Strings}, this reduction is still limited. Nevertheless, the homogeneous equation \eqref{Eq: Oc homogeneous eq} can be solve analytically for small system sizes via symbolic computation in Mathematica, which might bring the insight to the construction of zero mode to generic models. In the algorithm, \eqref{Eq: Oc homogeneous eq} is first simplified by Gaussian elimination, which leads to constraints on the coefficients $\{\alpha\}$. Since we are only interested in solutions with non-zero overlap with $\sigma_1^x$, we require that $\alpha_1 \neq 0$ in \eqref{Eq: Oc in terms of Pauli Strings}, and we denote the $n$ linearly independent solutions as $\{O_{c;1}, O_{c;2}, \ldots, O_{c,n}\}$. Note that these $n$ linearly independent solutions are not orthonormal. One first normalizes $\{O_{c;1}, O_{c;2}, \ldots, O_{c,n}\}$ to obtain $\{o_{c;1}, o_{c;2}, \ldots, o_{c;n}\}$ and then performs a Gram-Schmidt algorithm to obtain orthonormal eigenvectors $\{|\lambda_1), |\lambda_2), \ldots, |\lambda_n)\}$. Initially, we set $|\lambda_1) = |o_{c;1})$. Iteration for $j > 1$ follows
\begin{align}
    &|\lambda_j^\prime) = |o_{c;j}) - \sum_{i=1}^{j-1}|\lambda_i)(\lambda_i|o_{c;j}),\\
    &|\lambda_j) = \frac{|\lambda_j^\prime)}{\sqrt{(\lambda_j^\prime|\lambda_j^\prime)}},
\end{align}
where the inner product is defined as $(A|B) = \text{Tr}[A^\dagger B]/2^L$. Note that the distinct strings of spin operators form an orthonormal basis with respect to this inner product. Also, the number $n$ of orthonormal eigenvectors could be smaller than the size of the zero eigenvalue sector $N$ in \eqref{Eq: Oc projected to zero sector} since only the eigenvectors with non-zero overlap with $\sigma_1^x$ are included in the algorithm. Finally, one may construct $O_c$ based on \eqref{Eq: Oc projected to zero sector}.

In the following, we present explicit expressions for the linearly independent solutions $\{ O_{c;j}^{A;L} \}$ and $\{ O_{c;j}^{B;L} \}$ of $\Tilde{H}_L^A$ and $\Tilde{H}_L^B$ for small system sizes. For $L=3$, $\tilde{H}^A_3=\tilde{H}^B_3$ and therefore $O_{c;j}^{A;3} = O_{c;j}^{B;3}$. There are two independent solutions,
\begin{align}
    O_{c;1}^{L=3} = &\sigma^x_1+g\sigma^z_1\sigma^x_2+g^2\sigma^z_1\sigma^z_2\sigma^x_3-J_z\sigma^y_1\sigma^y_2\sigma^x_3\nonumber\\
    & +J_z g(\sigma^z_1+\sigma^z_2)\sigma^x_3,\label{eq:ol3c1}\\
    O_{c;2}^{L=3} = & g^2\sigma^x_1-g J_z\sigma^x_1\sigma^z_2+g(g^2-J_z^2)\sigma^z_1\sigma^x_2\nonumber\\
    &+gJ_z(1+g^2-J_z^2)\sigma^z_2\sigma^x_3+J_z(g^2-J_z^2)\sigma^x_1\sigma^x_2\sigma^x_3\nonumber\\
    &+gJ_z(g^2-J_z^2)\sigma^z_1\sigma^x_3\nonumber\\
    &+g^2(g^2-J_z^2)\sigma^z_1\sigma^z_2\sigma^x_3,\label{eq:ol3c2}
\end{align}
where the superscript $A$ and $B$ is omitted. Interestingly, these two operators are distinct for $J_z>0$ but reduce to the zero mode of the TFIM \eqref{Eq: Zero mode of Transverse-field Ising} as $J_z\to 0$,
\begin{equation}
    O_{c;1,2}^{L=3}\to \psi_0 \propto \sigma^x_1+g\sigma^z_1\sigma^x_2+g^2\sigma^z_1\sigma^z_2\sigma^x_3.
\end{equation}
A particular consequence is that, after normalization, the overlap of $O_{c;1,2}^{L=3}$ with $\sigma^x_1$ is of $\mathcal{O}(1)$ in $J_z$. By expressing $O_{c;1,2}^{L=3}$ in terms of Majoranas, the interactions lead to the three-Majorana terms (terms with $J_z$).
\begin{align}
    O_{c;1}^{L=3}=&a_1+ga_3+g^2a_5-iJ_za_2a_3a_5-igJ_za_3a_4a_5\nonumber\\
    &-igJ_za_1a_2a_5,\label{eq:L31}\\
    O_{c;2}^{L=3}=&g^2a_1+g(g^2 - J_z^2)a_3+g^2(g^2 - J_z^2)a_5\nonumber\\
    &+igJ_za_1a_3a_4-iJ_z(g^2 - J_z^2)a_1a_4a_5\nonumber\\
    &-igJ_z(g^2 - J_z^2)a_3a_4a_5\nonumber\\
    &-igJ_z(1 + g^2 - J_z^2)a_1a_2a_5.\label{eq:L32}
\end{align}

For chain length $L=4$, $\tilde{H}^A_4\neq \tilde{H}^B_4$, and indeed $O^A_c$ and $O^B_c$ are different. For the impurity model $\tilde{H}^A_4$, there are four linearly independent solutions $O^{A;L=4}_{c;1,...,4}$ with non-zero overlap with $\sigma^x_1$, but the overlap vanishes as $J_z \rightarrow 0$. 
However, this seems to be an artifact of the $L=4$ case as we have checked different system sizes up to $L=7$. For the model with interactions on all sites $\tilde{H}^B_4$, there are five linearly independent solutions $O^{B;L=4}_{c;1,...,5}$ with non-zero overlap with $\sigma_1^x$. Of these, three of them have $O(1)$ overlap with $\sigma^x_1$ and the other two have  $O(J_z)$ overlap.
Explicitly, the simplest operator is
\begin{align}
    O^{B;L=4}_{c;1}=& 3a_1+3ga_3+3g^2a_5+3g(g^2 + J_z^2)a_7\nonumber\\
    &-iJ_za_1a_3a_6-3iJ_za_2a_3a_5-igJ_za_3a_6a_7\nonumber\\
    &-igJ_za_1a_4a_7-3igJ_za_3a_4a_5-4igJ_za_4a_5a_7\nonumber\\
    &-4igJ_za_2a_3a_7-4igJ_za_1a_2a_5\nonumber\\
    &-5ig^2J_za_5a_6a_7-6ig^2J_za_3a_4a_7\nonumber\\
    &+i(1 - 5g^2)J_za_1a_2a_7-ga_1a_3a_4a_6a_7\nonumber\\
    &-ga_1a_2a_3a_5a_6-g^2a_2a_3a_5a_6a_7-g^2a_1a_2a_4a_5a_7\nonumber\\
    &-g^2a_1a_2a_3a_4a_5+g(1 - g^2 - J_z^2)a_1a_2a_5a_6a_7\nonumber\\
    &+g(1 - g^2 - 4J_z^2)a_1a_2a_3a_4a_7\nonumber\\
    &-g(g^2 + 4J_z^2)a_3a_4a_5a_6a_7.\label{eq:L4}
\end{align}
For $J_z\to 0$, note that $O^{B;L=4}_{c;1}\to \psi_0+(\text{five Majorana terms})$, which indeed gives an $\mathcal{O}(1)$ overlap with $a_1=\sigma^x_1$.

\begin{figure}[h!]
    \centering \includegraphics[width=0.4\textwidth]{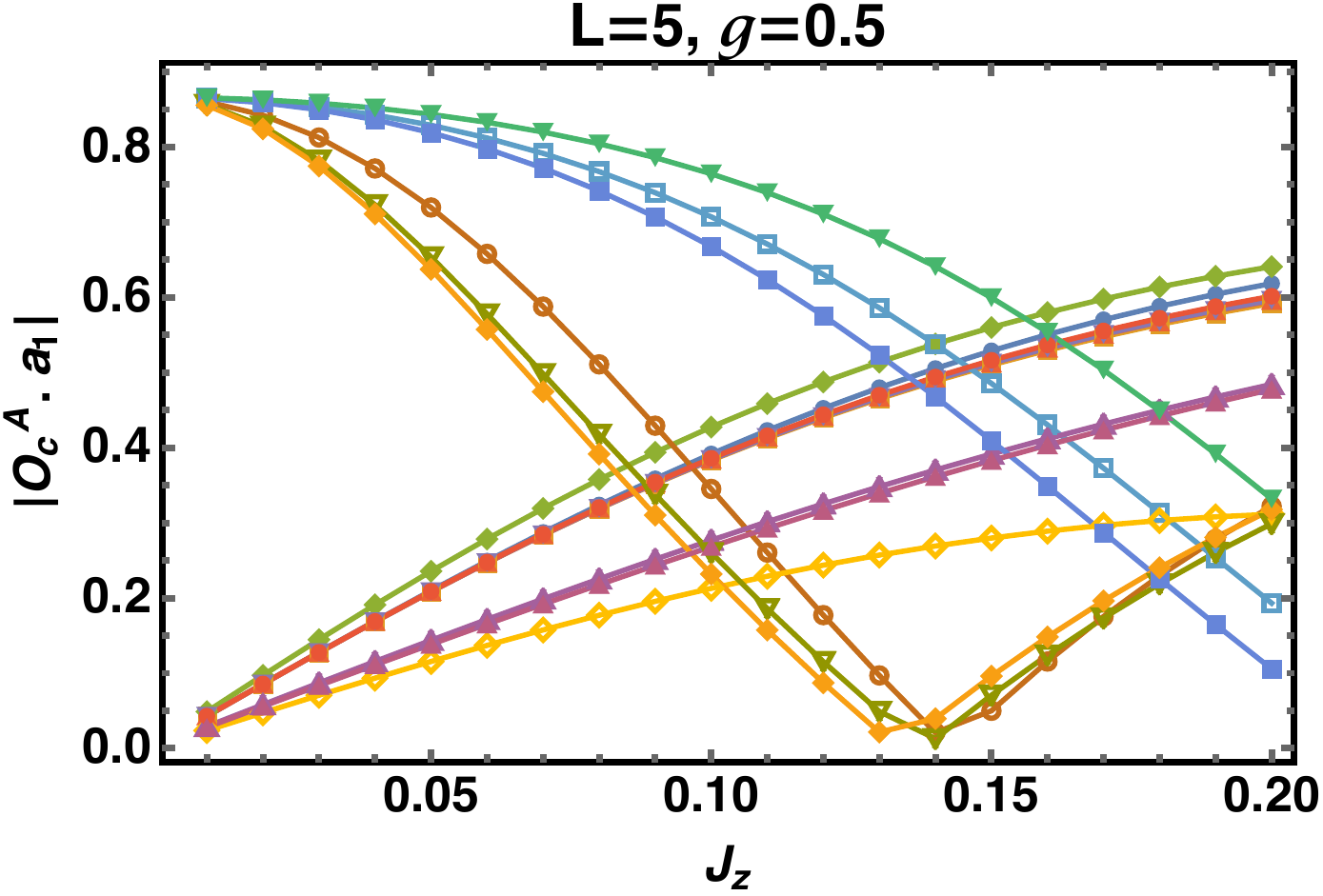} \includegraphics[width=0.4\textwidth]{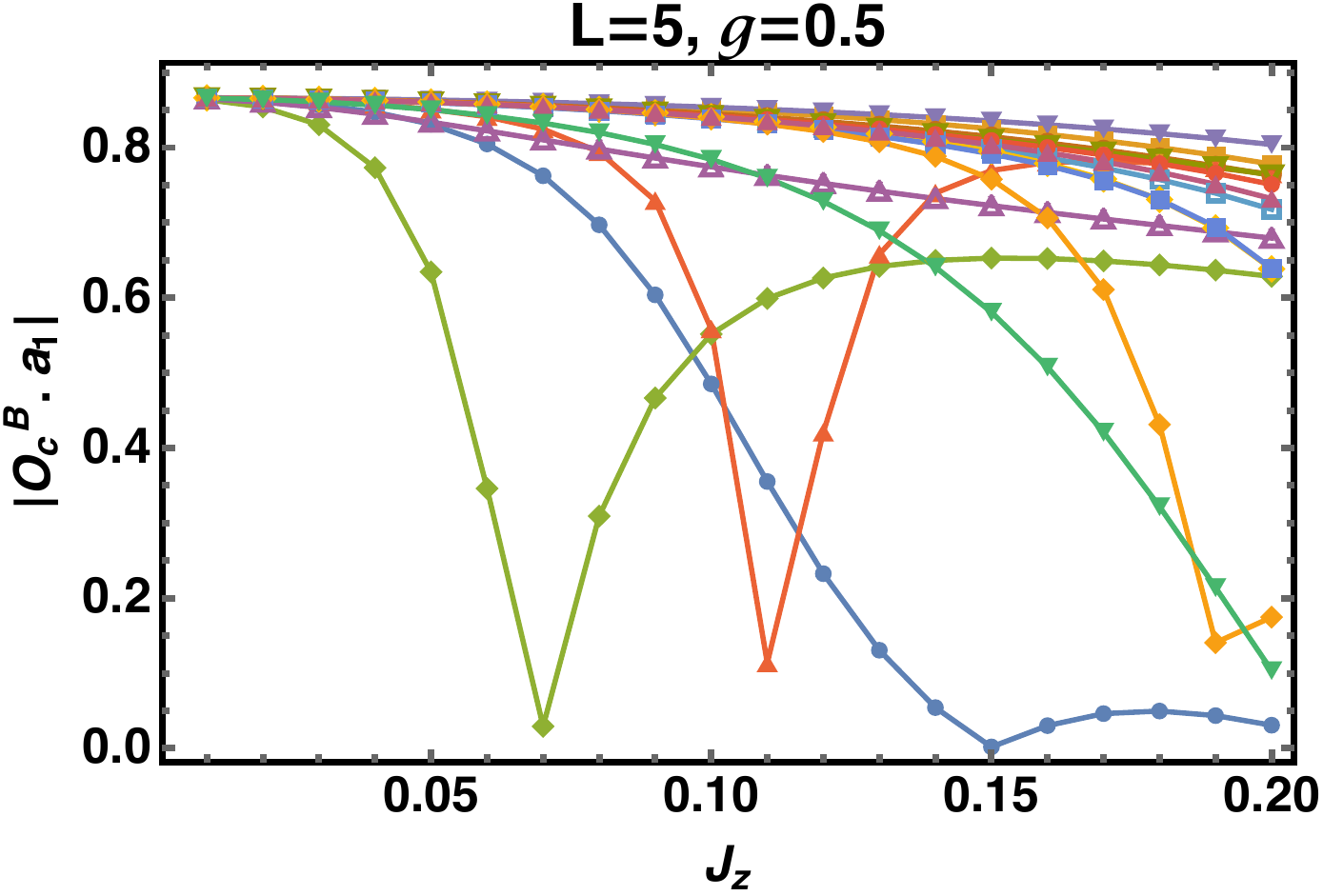}
    \caption{A basis of the conserved operators $O^{A,B}_c$  which overlap with $a_1=\sigma^x_1$ for size $L=5$. We plot the overlap as a function of $J_z$. We see that for both the impurity model $\tilde{H}^A_5$ (top panel) and the model with interactions on all sites $\tilde{H}^B_5$ (bottom panel) there are conserved quantities with an overlap of order $1$ for small $J_z$. For the impurity model there are also conserved quantities with overlap of order $J_z$.} 
    \label{fig: L5 conserved quantities}
\end{figure}
For $L=5,6,7$, one can also find conserved quantities $O_c^{A,B;L}$, with overlap of $\mathcal{O}(1)$ with $\sigma_1^x$, i.e., $a_1$. Since there are too many solutions of the quasi-conserved operators, we simply show the overlap of a full set of linearly independent $O_{c;i}^{A,B;L}$ with $a_1$ for $L=5$ in Fig.~\ref{fig: L5 conserved quantities}.

\begin{figure}[h!]
    \centering \includegraphics[width=0.45\textwidth]{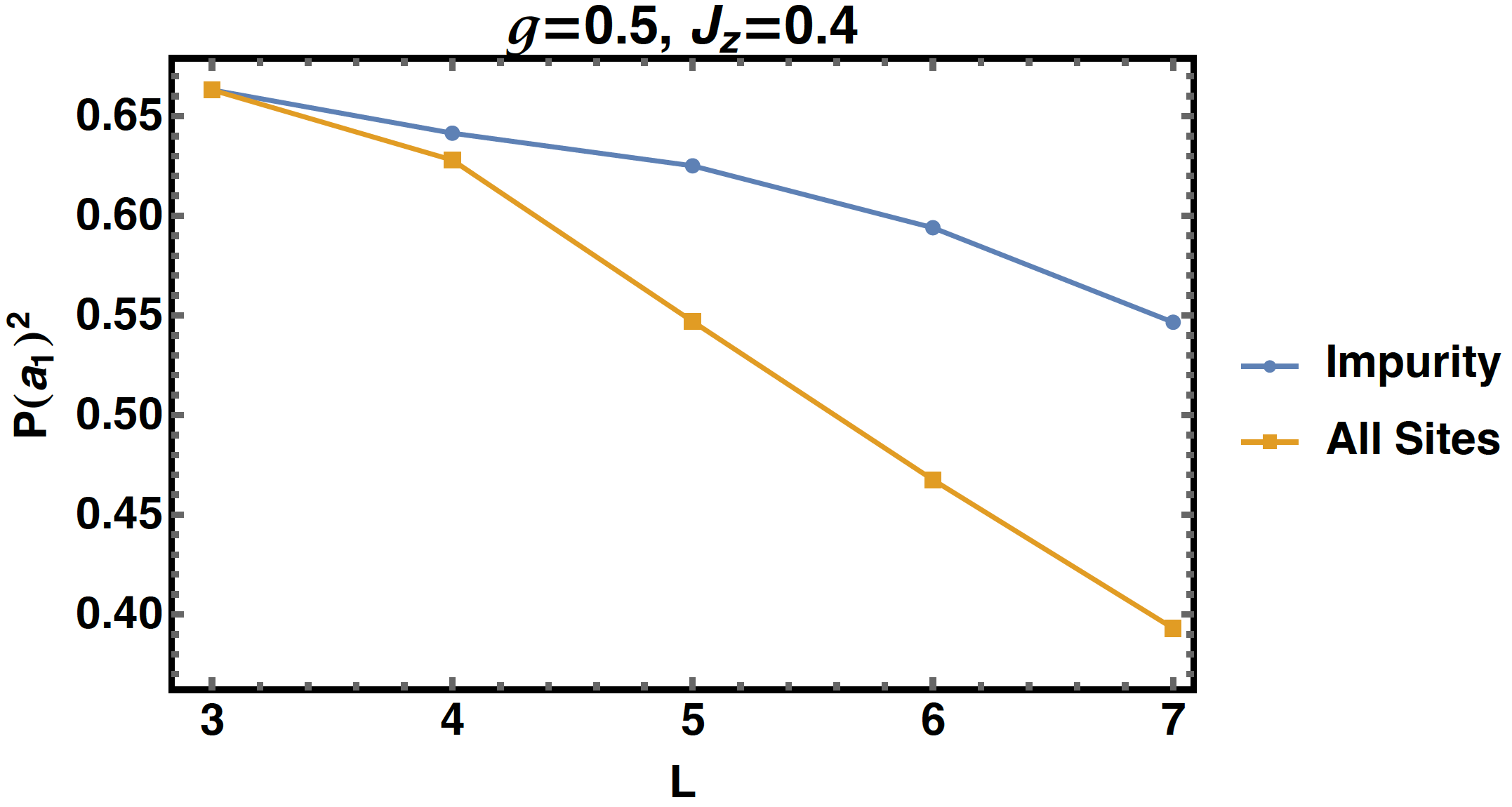} \includegraphics[width=0.45\textwidth]{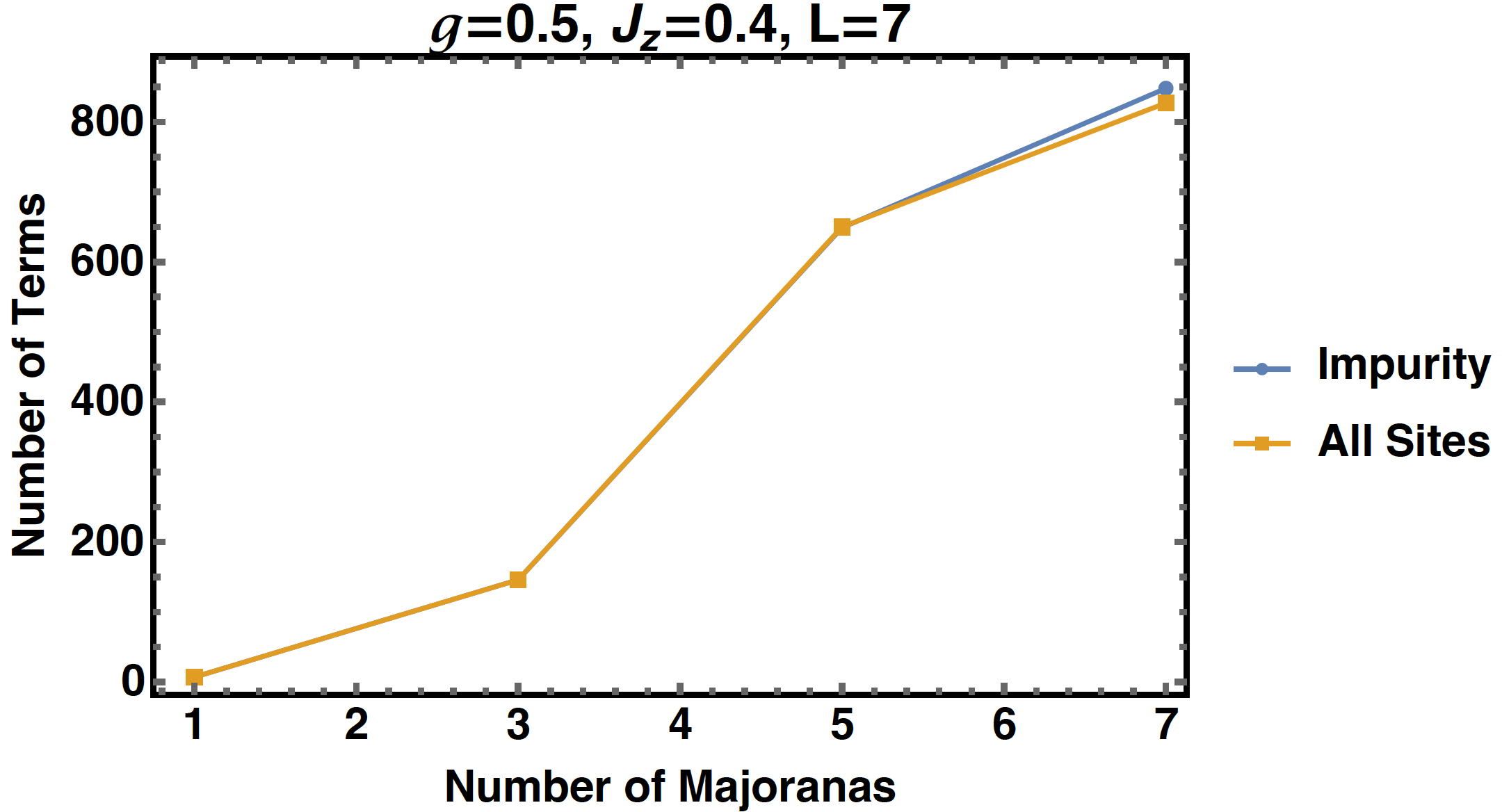}
    \caption{Top panel: Squared norm of the projection of the edge spin $\sigma^x_1=a_1$ onto the subspace of operators spanned by {\color{blue} $\{|\lambda_j)\}$}. For both models, this projection becomes smaller for a larger chain, signaling that the resulting plateau observed in the autocorrelation function becomes smaller. Bottom panel: Number of terms as a function of string length in the conserved operator $O^{A,B;L=7}_c$  with largest overlap with the edge spin. The rapid increase in the number of terms is responsible for the delocalization of the conserved quantities.} 
    \label{fig: delocalized}
\end{figure}

The square norm of $a_1$ projected on the vector space spanned by $\{ |\lambda_j) \}$ is defined as
\begin{align}
    P(a_1)^2=\sum_{j=1}^n|(\lambda_j|a_1)|^2.
\end{align}
This accounts for the value of the late time plateau of the autocorrelation function according to \eqref{Eq: Oc projected to zero sector}.  
As we show in Fig.~\ref{fig: delocalized}, the projected norm decreases with the size of the chain $L$, and is expected to approach the values of the plateaus in Fig.~\ref{fig: autocorrelation and conserved case} upon extrapolation to $L=10$. We can understand the decrease of the norm of this operator as due to its delocalization. As we noted in equations (\ref{eq:L31}-\ref{eq:L4}), for a longer chain the conserved operators $O_c$ involve longer Majorana strings and the number of possible strings increases exponentially. Fig.~\ref{fig: delocalized} shows that the number of terms involving longer strings of Majoranas indeed increases very rapidly, which leads to the $O_c$ becoming less localized on the first site. Note also that, while the number of terms increases at the same rate for both the impurity model and the model with interactions on all sites, in the impurity model the weight of the longer strings is smaller, which is consistent with this model displaying higher plateaus (Fig. \ref{fig: autocorrelation and conserved case}).

\section{Random state approximation and Trotter decomposition} \label{Appendix:Random state and Trotter}

Due to the limitations of ED, a different numerical method is needed in order to explore autocorrelation functions for large system sizes. The autocorrelation of $\sigma_1^x$ \eqref{Eq: AutoCorrelation Sigmax} is explicitly written as 
\begin{align}
    A_\infty (t) = \frac{1}{2^L} \text{Tr}\left[ U^\dagger(t)\sigma_1^x U(t) \sigma_1^x \right],
\end{align}
where $U(t)$ is the unitary evolution operator. One can replace the last $\sigma_1^x$ by $(\sigma_1^x+\mathbb{I})$, where $\mathbb{I}$ is the identity matrix, since $\text{Tr}[\sigma_1^x] = \text{Tr}[\sigma_1^x(t)] = 0$ so that the autocorrelation stays the same.  Moreover, with $(\sigma_1^x)^2 = \mathbb{I}$, one derives the identity $(\sigma_1^x+\mathbb{I}) = (\sigma_1^x+\mathbb{I})^2/2$. By cyclic permutation in the trace, the autocorrelation function has the following symmetric form
\begin{align}
   A_\infty (t) = \frac{1}{2^L}\text{Tr}\left[\frac{ \left( \sigma_1^x+\mathbb{I}\right)}{\sqrt{2}}U^\dagger (t) \sigma_1^x U(t)\frac{ \left(\sigma_1^x+\mathbb{I}\right)}{\sqrt{2}} \right].
\end{align}
Now, we approximate the trace by average over a Haar random state $|\phi\rangle$ up to $\mathcal{O}(1/\sqrt{2^L})$ corrections
\begin{align}
    A_\infty (t) \approx \left\langle \phi \left| \frac{ \left( \sigma_1^x+\mathbb{I}\right)}{\sqrt{2}}U^\dagger (t) \sigma_1^x U(t)\frac{ \left(\sigma_1^x+\mathbb{I}\right)}{\sqrt{2}} \right| \phi\right\rangle.
    \label{Eq: Haar random approx}
\end{align} 
This approximation can be justified by the following argument. For a Haar random state expanded in eigenstate bases, $|\phi\rangle = \sum_{n=1}^{2^L} c_n|n\rangle$, typically each coefficient $c_n$ has size $1/\sqrt{2^L}$ with a random phase. For a given matrix $M$, 
the average over a Haar random state is
\begin{align}
    \langle \phi | M | \phi \rangle = \sum_{n=1}^{2^L} |c_n|^2 \langle n | M | n \rangle + \sum_{\substack{n, m = 1 \\ n\neq m}}^{2^L} c_n^* c_m  \langle n | M | m \rangle,
\end{align}
where the first term leads to $\text{Tr}[M]/2^L$ since $|c_n|^2 \sim 1/2^L$. The difference between Haar random state average and the trace comes from the second term. To estimate the size of the second term, we take the square of it
\begin{align}
    &\sum_{\substack{n, m = 1 \\ n\neq m}}^{2^L} \sum_{\substack{k, l = 1 \\ k\neq l}}^{2^L} c_n^* c_m c_k^* c_l  \langle n | M | m \rangle    \langle k | M | l \rangle \nonumber\\
    &= \sum_{\substack{n, m = 1 \\ n\neq m}}^{2^L} |c_n|^2 |c_m|^2 |\langle n | M | m \rangle|^2 \sim \frac{1}{2^L} \cdot  \frac{1}{2^L}\text{Tr}[M^\dagger M].
\end{align}
Due to the randomness of the coefficients, only the terms with $n = l$ and $m = k$ survive in the summation. In the last line, $|c_n|^2 \sim |c_m|^2 \sim 1/2^L$ and the identity $\sum_{n,m} |\langle n | M | m \rangle|^2 = \text{Tr}[M^\dagger M]$ are used. Although the identity is only true when the summation includes $n = m$ terms, it does not matter here since one only needs to estimate the order of magnitude of this summation. In this article, we focus on $M = \sigma_1^x(t)\sigma_1^x$ and  $\text{Tr}[M^\dagger M]/2^L = \mathcal{O}(1)$. Therefore, the Haar random state average gives a good approximation of the trace upto $\mathcal{O}(1/\sqrt{2^L})$ corrections as we claim in \eqref{Eq: Haar random approx}. 

Based on \eqref{Eq: Haar random approx}, we define a new time-evolving state, $|\Tilde{\phi}(t)\rangle = U(t)[(\sigma_1^x+\mathbb{I})/\sqrt{2}]|\phi\rangle$, and the autocorrelation becomes
\begin{align}
    A_\infty (t) \approx \langle \Tilde{\phi}(t)| \sigma_1^x |\Tilde{\phi}(t)\rangle.
\end{align}
This representation of the autocorrelation function has advantages for large system sizes. It costs much less memory resources to evolve a state with $2^L$ components than performing ED on a $2^L\times2^L$ matrix. However, the computation time depends linearly on $t$ as the number of time steps to evolve $|\Tilde{\phi}\rangle$ to  $|\Tilde{\phi}(t)\rangle$ is proportional to $t$.

For a unitary evolution in time step $dt$, we apply Trotter-decomposition,
\begin{align}
    U(dt) \approx  e^{-iH_{\rm xx}dt}e^{-i H_z dt}e^{-i H_{\rm zz} dt},
\end{align}
where $H_{\rm xx}, H_z$ and $H_{\rm zz}$ correspond to the three terms in the Hamiltonian \eqref{Eq: Impurity Hamiltonian}. The many-body state is represented in the $\sigma^z$ basis so that $e^{-i H_z dt}e^{-i H_{zz} dt}$ is diagonal. The nearest neighbor interaction terms in $H_{\rm xx}$ commute with each other so that $ e^{-iH_{\rm xx}dt}$ is a series product of nearest neighbor unitary evolution. The unitary evolution in one time step is explicitly expressed as
\begin{align}
    &U(dt) \nonumber\\
    &\approx \left[\prod_{j=1}^{L-1} \biggl\{ \cos{(dt) - i\sin{(dt)\sigma_j^x\sigma_{j+1}^x}} \biggr\}\right]e^{-i H_z dt}e^{-i H_{\rm zz} dt},
\end{align}
where $e^{-i H_z dt}e^{-i H_{\rm zz} dt}$ is a diagonal matrix with $2^L$ non-zero elements and $\sigma_j^x\sigma_{j+1}^x$ permutes different many-body states and is a sparse matrix with $2^L$ non-zero elements. Thus these objects are efficient in memory resources, but the overall computation time increases with system size. 

\begin{figure}[h!]
    \centering \includegraphics[width=0.45\textwidth]{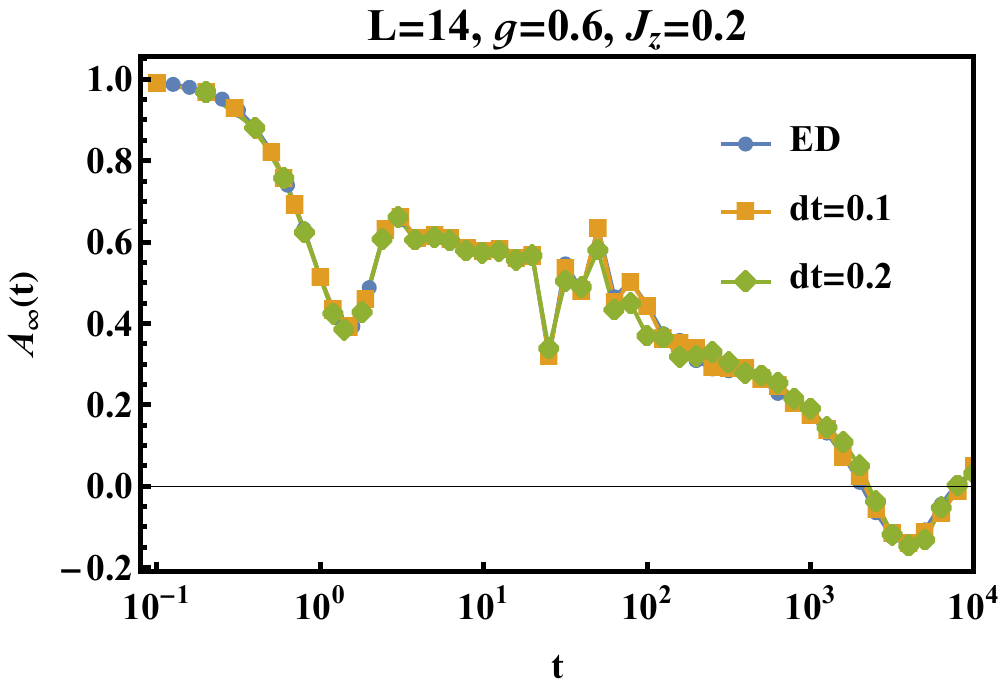} \includegraphics[width=0.45\textwidth]{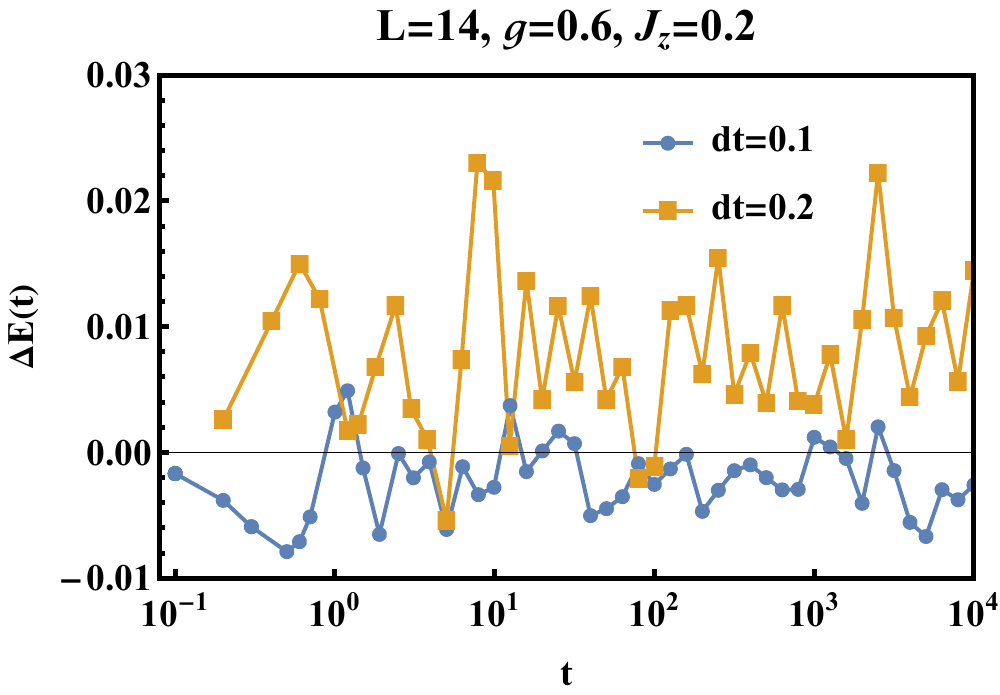}
    \caption{Top panel: Autocorrelation function calculated by ED and by random state average with Trotter decomposition $dt=0.1,0.2$. For $dt = 0.1$, the approximate results agree well with ED. With $dt = 0.2$, there are slight deviations from ED but still the key features of ED are captured. Bottom panel: Energy fluctuations for the random state average with Trotter decomposition. The energy difference is measured from the initial energy at $t=0$. The fluctuations are larger for larger time step. Both $dt = 0.1,0.2$ do not show a steady heating, consistent with a high frequency driving related to $2\pi/dt \gg 1$.}
    \label{fig: ED vs RandomTrotter}
\end{figure}

\begin{figure}[h!]
    \centering \includegraphics[width=0.45\textwidth]{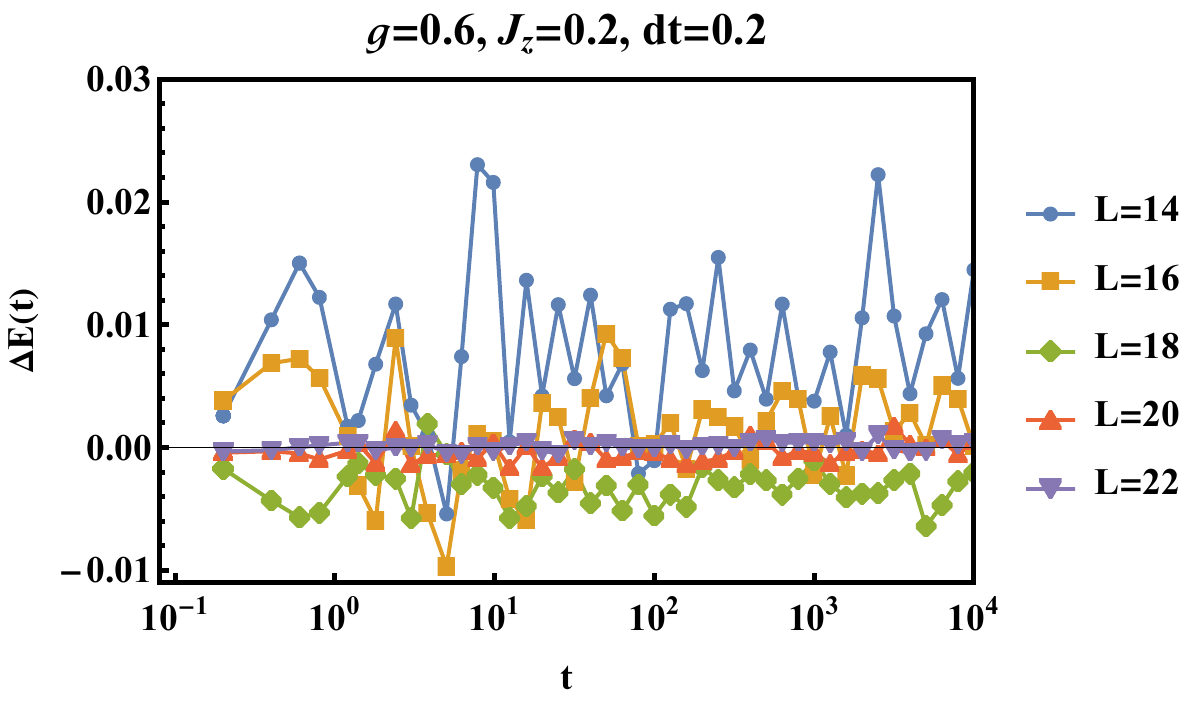}
    \caption{Energy fluctuations in the results of the random state average with Trotter decomposition $dt=0.2$, and for different system sizes. The energy difference is measured from the initial energy at $t=0$. As the system size increases, the energy fluctuations decrease.}
    \label{fig: energy}
\end{figure}

Fig.~\ref{fig: ED vs RandomTrotter} shows the comparison between ED and the approximate method just described, with time steps $dt = 0.1,0.2$ (top panel). $dt = 0.1$ is consistent with ED results. However, to reduce the computation time, we take $dt = 0.2$ in the main text such that the key features of the autocorrelation function are still captured. We sacrifice some precision in order to explore larger system sizes. The energy fluctuation in the bottom panel of Fig.~\ref{fig: ED vs RandomTrotter} confirms that the system is not heating because of the high-frequency drive (small time step). The fluctuations become smaller for smaller time steps as one expects energy conservation to be recovered in the continuous-time limit. Fig.~\ref{fig: energy} shows the energy fluctuations for different system sizes. The fluctuations are suppressed for larger system sizes.

\section{Fermi's Golden Rule}
\label{Appendix:FGR}

We present the full derivation of the FGR decay rate of the infinite temperature autocorrelation of the zero modes. The full Hamiltonian consists of two parts: the perturbing interaction $V$ and the unperturbed Hamiltonian $H_0$. The unitary evolution up to time $t$ is 
\begin{align}
    U(t) &= e^{-i (H_0+V) t}.
\end{align}
The time evolution of the zero modes up to time $t$, $\psi_0(t)=U(t)^\dagger \psi_0 U(t)$, can be expressed as
\begin{align}
    \psi_0(t) =  e^{i (\mathcal{L}_0+\mathcal{L}_V) t}\psi_0,
\end{align}
where the notations are as follows: $\mathcal{L}_0 \psi_0 = [H_0, \psi_0]$ and $\mathcal{L}_V \psi_0 = [V,\psi_0]$. The infinite temperature autocorrelation is given by
\begin{align}
    A_\infty (t) = \frac{1}{2^L} \text{Tr}[\psi_0(t)\psi_0].
\end{align}
We will only expand up to second order in $V$ and denote $A_{\infty,n}$ as the autocorrelation function to $n$-th order in $V$. The time order expansion of  $\psi_0(t)$ up to second order in $\mathcal{L}_V$ is
\begin{align}
    &\psi_0(t)\nonumber\\
    &\approx e^{i \mathcal{L}_0 t}\psi_0 + \int_0^t dt' e^{i \mathcal{L}_0 (t-t')}(i\mathcal{L}_V)e^{i \mathcal{L}_0 t'}\psi_0 \nonumber\\
    & + \int_0^t dt'' \int_{t''}^t dt'  e^{i \mathcal{L}_0 (t-t')}(i\mathcal{L}_V)e^{i \mathcal{L}_0 (t'-t'')}(i\mathcal{L}_V)e^{i \mathcal{L}_0 t''}\psi_0.
\end{align}
At the zeroth order, one does not pick up any terms containing $\mathcal{L}_V$, so that 
\begin{align}
    A_{\infty,0}(t) = \frac{1}{2^L} \text{Tr}\left[\left\{ e^{i\mathcal{L}_0 t} \psi_0 \right\}\psi_0 \right] = 1,
\end{align}
where we have used the commutation relation of the zero mode $\mathcal{L}_0\psi_0 = 0$ and also employed the normalization $\text{Tr}[\psi_0 \psi_0]/2^L = 1$. 
Note that while $\mathcal{L}_0\psi_0$ is not exactly zero for a finite system,  it is exponentially small in system size and negligible in the computation of decay rate.

At first order,  $\mathcal{L}_V$ appears once in the expansion
\begin{align}
    &A_{\infty,1}(t) \nonumber\\ 
    &= \frac{1}{2^L}\int_0^t dt'\text{Tr}\left[ \left\{e^{i \mathcal{L}_0 (t-t')}(i\mathcal{L}_V)e^{i \mathcal{L}_0 t'}\psi_0\right\} \psi_0\right].
\end{align}
With cyclic permutation within the trace, one can show that $\text{Tr}\left[ \left\{e^{i\mathcal{L}_0t}  O_1\right\} O_2\right] = \text{Tr}\left[ O_1 \left\{e^{-i\mathcal{L}_0t}  O_2\right\} \right]$ for arbitrary operators $O_1$ and $O_2$. Also, from the commutation relations,  the first-order expansion is further simplified as
\begin{align}
    A_{\infty,1}(t) = \frac{1}{2^L}\int_0^t dt'\text{Tr}\left[ \left\{ (i\mathcal{L}_V) \psi_0\right\} \psi_0\right] = 0,
\end{align}
which is traceless due to the cyclic property of trace: $\text{Tr}\left[ \left\{ \mathcal{L}_V \psi_0\right\} \psi_0\right] = \text{Tr}\left[ \psi_0 \left\{ -\mathcal{L}_V \psi_0\right\} \right] = 0$.

\begin{figure}
    \centering \includegraphics[width=0.45\textwidth]{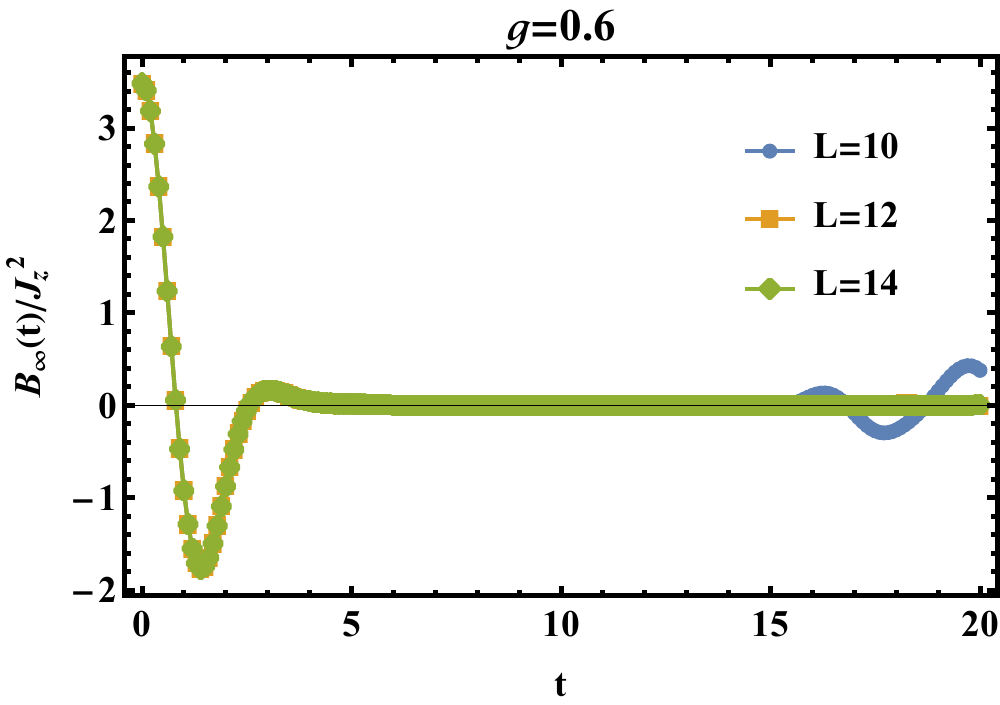} \includegraphics[width=0.45\textwidth]{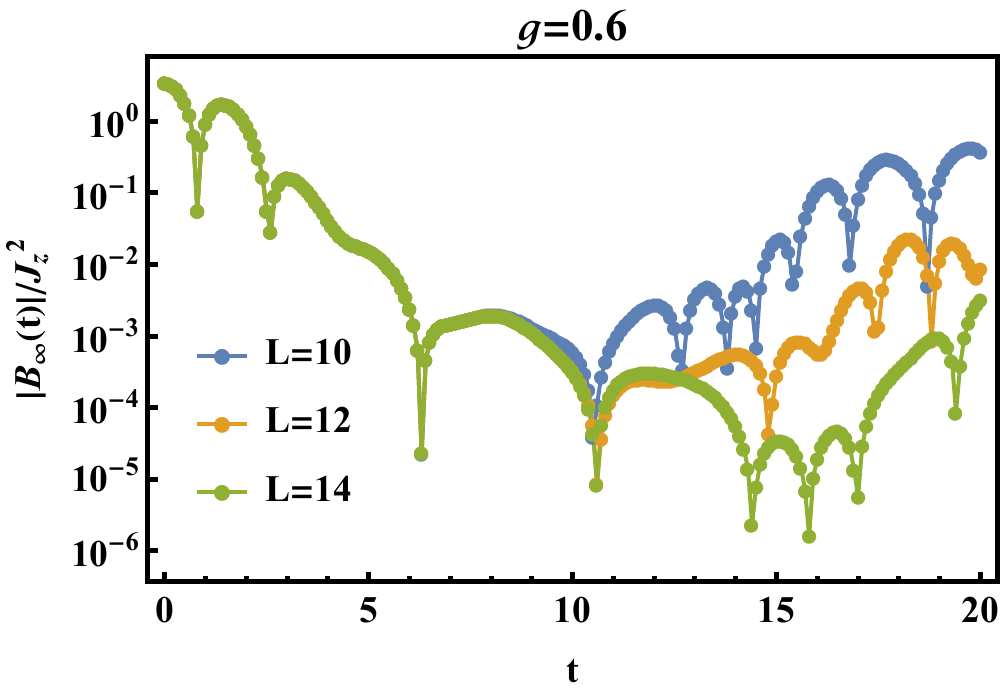}
    \caption{The infinite temperature autocorrelation of $\Dot{\psi_0}$,  $B_{\infty}=\text{Tr}[\dot{\psi_0}(t) \dot{\psi_0}]/ 2^L$. $J_z$ is the boundary impurity strength and the results are $J_z$-independent after multiplying by $1/J_z^2$. The top panel shows a fast decay that supports the approximation in ~\eqref{Eq: 2nd Order of A}. The fluctuations become large at a later time for $L=10$, which is the revival effect of a finite-size system. This can be clearly seen in the bottom panel, where the revivals move to later times as the system size increases. To take finite system size into account, we truncate the integral of $t$ in \eqref{Eq: Fermi's Golden Rule 0Mode} up to the minimum value in the bottom panel. We obtain the decay rate $\Gamma = 0.16J_z^2$ for $g = 0.6$.}
    \label{fig: B autocorrelation}
\end{figure}

Finally, for the second order correction,  $\mathcal{L}_V$ appears twice in the expansion
\begin{align}
    &A_{\infty,2}(t) \nonumber\\
    &= \frac{1}{2^L} \int_0^t dt'' \int_{t''}^t dt' \text{Tr}\left[\left\{e^{i \mathcal{L}_0 (t-t')}(i\mathcal{L}_V)e^{i \mathcal{L}_0 (t'-t'')} \right. \right. \nonumber\\
    &\qquad\qquad\qquad\qquad\qquad 
    \left. \left. \times(i\mathcal{L}_V)e^{i \mathcal{L}_0 t''}\psi_0\right\} \psi_0\right].
    \label{Eq: 2nd order expansion}
\end{align}
As we have learnt from the first order expansion, $e^{i\mathcal{L}_0 (t-t')}$ and $e^{i\mathcal{L}_0 t''}$ contribute an overall factor 1. Then, one associates the first $i\mathcal{L}_V$ with the last $\psi_0$ by cyclic permutation. One obtains
\begin{align}
   &A_{\infty,2}(t) \nonumber\\
   &= -\frac{1}{2^L} \int_0^t dt'' \int_{t''}^t dt' \text{Tr}\left[\dot{\psi_0}(t'-t'') \dot{\psi_0}\right] \nonumber\\
   &= -\frac{1}{2^L} \int_0^t dt'' \int_0^t dt' \theta(t-t'-t'') \text{Tr}\left[\dot{\psi_0}(t') \dot{\psi_0}\right] \nonumber\\
   &= -\frac{t}{2^L}\int_0^t dt'  \left( 1 - \frac{t'}{t}\right)\text{Tr}\left[\dot{\psi_0}(t') \dot{\psi_0}\right],
\end{align}
where we define $\dot{\psi_0} = i\mathcal{L}_V \psi_0$ and $\dot{\psi_0}(t) = e^{i\mathcal{L}_0 t}\dot{\psi_0}$. In the third line, we shift $t' \rightarrow t'+t''$ and impose the Heaviside theta function to preserve time order.

On combining the above results, the autocorrelation function up to second order in $V$ is
\begin{align}
    A_\infty (t) \approx A_{\infty,0}(t) + A_{\infty,2}(t),
\end{align}
where
\begin{align}
    &A_{\infty,0} (t) = 1,\\ 
    &A_{\infty,2} (t) = -\frac{t}{2^L}\int_0^\infty dt' \text{Tr}\left[\dot{\psi_0}(t') \dot{\psi_0}\right]  \label{Eq: 2nd Order of A}
\end{align}
Note that we approximate the upper bound of the integral $t$ by $\infty$, and therefore the $(1-t'/t)$ in the summation is replaced by 1. Since we study quantities where the lifetime is long, $t$ is chosen to be a large number. In addition, $\text{Tr}[\dot{\psi_0}(t') \dot{\psi_0}]$ decays fast with a time scale that is much smaller than $t$. Therefore, we can simply replace $t$ by $\infty$ in the integral.

The autocorrelation function with decay rate $\Gamma$ can be formulated as $A_\infty (t) = e^{-\Gamma t} \approx (1-\Gamma t)$. By comparing this to the second-order expansion, we obtain the FGR decay rate
\begin{align}
    \Gamma = \frac{1}{2^L} \int_0^{\infty} dt\ \text{Tr}[\Dot{\psi}_0(t)\Dot{\psi}_0(0)],
\end{align}
which is \eqref{Eq: Fermi's Golden Rule 0Mode} in the main text.

Fig.~\ref{fig: B autocorrelation} demonstrates the numerical computation of the infinite temperature autocorrelation $B_\infty(t) = \text{Tr}[\dot{\psi_0}(t) \dot{\psi_0}]/ 2^L$,  and the decay rate derived from it based on \eqref{Eq: Fermi's Golden Rule 0Mode}. The top panel validates the approximation in \eqref{Eq: 2nd Order of A} where $(1-t'/t)$ is replaced by 1. The numerical time integral is truncated at the minimum in the bottom panel to account for finite system size effects.


%

\end{document}